\documentclass[a4paper,fleqn,usenatbib,useAMS]{mnras}

\usepackage{graphicx}	
\usepackage{amsmath}	
\usepackage{amssymb}	
\usepackage{multicol}	
\usepackage{bm}			
\usepackage{pdflscape}	
\usepackage[T1]{fontenc}
\usepackage{ae,aecompl}
\usepackage{newtxtext, newtxmath}

\newcommand{\added}[1]{\textcolor{black}{#1}}

\title[Long-term evolution of SMBHBs]
{Non-steady-state long-term evolution of supermassive black hole binaries surrounded by accretion discs}
\author[C. Fontecilla, Z. Haiman \& J. Cuadra]{
{Camilo Fontecilla}$^{1}$\thanks{E-mail: cfonteci@astro.puc.cl}, {Zolt\'an Haiman}$^{2}$ and {Jorge Cuadra}$^{1}$
\\
$^{1}$ Instituto de Astrof\'isica, Pontificia Universidad Cat\'olica de Chile, Av. Vicu\~na Mackenna 4860, Santiago, Chile\\
$^{2}$ Department of Astronomy, Columbia University, New York, NY 10027, USA\\
}

\date{Accepted 20XX . Received 20XX}
\pubyear{20XX}

\begin{document}
\label{firstpage}
\pagerange{\pageref{firstpage}--\pageref{lastpage}} 
\maketitle

\begin{abstract}
Supermassive black holes (SMBHs) pair and form bound binaries after their host galaxies merge. In a gas-rich merger, accretion discs are expected to form around the binary and its components. These discs control the binary orbital evolution until the system is compact enough for gravitational waves to drive the SMBHs to coalescence.
In this work, we implemented a time-dependent 1D model to follow the long-term evolution of the coupled binary + discs system, from a separation of $10^5$ down to 20 Schwarzschild radii.
We run different models changing the system parameters, including the binary mass ratio $q \leq 0.3$ and a factor $\gamma$ that controls the inflow across the gap created by the secondary. 
We find that our implementation yields higher residual masses and longer binary residence times than previous studies.
Our main conclusion is the non-steady-state nature of the evolution of the system: the properties the disc had when the binary was still at large separations influence its whole evolution. To recover steady state, the binary residence time would have to be much longer than the inflow time-scale of the disc throughout their entire history, which in general is not satisfied.
\end{abstract}


\section{Introduction} \label{sec:intro}

Observational evidence suggests the existence of supermassive black holes (SMBHs) at the centers of most, if not all, galaxies \citep[e.g.,][]{kormendy13}. Galaxies interact and merge according to the current hierarchical cosmological model, thus an expected outcome of those events is the formation of SMBH binaries in the galaxy remnants \citep[e.g.,][]{begelman80, volonteri03}. To become a bound binary and finally coalesce, the SMBHs will go through different processes that extract their energy and angular momentum, making them fall deeper into the gravitational potential \citep{begelman80}.

Coalescence can be achieved within a Hubble time \citep[depending of the environment conditions,][]{merritt05, colpi09, mayer18}, but the details of the mechanisms involved are still a topic of active research. While the galaxies are merging, dynamical friction between the individual SMBHs and the background of stars and gas will drag them into the center of the gravitational potential \citep{governato94}. After a bound binary is formed, dynamical friction becomes inefficient and the shrinkage will be dominated by scattering of individual stars, a process which efficiency depends on the properties of the stellar background \citep[e.g.,][]{yu02, vasiliev15}. In the gas-rich case, the gas that followed the black holes through the dynamical friction phase could form a circumbinary accretion disc around the binary \citep[e.g.,][]{escala05} and allow the system to shrink by angular momentum transfer due to a tidal interaction~\citep[e.g.][]{cuadra09,roedig11,tang17}. When the SMBHs reach a separation of~O(100) Schwarzchild radii (${\rm R_S}$), the emission of gravitational waves (GW) dominates and shrinks the binary in a short period of time \citep{peters64, armitage02, chang10}. Due to the different time-scales involved, most of the (accreting) SMBH binaries are expected to be in the accretion-disc dominated phase, which can take of the order of millions of years for the expected binary parameters \citep{haiman09, tang17}. Candidates for this kind of system have recently been identified as periodic quasars in the Catalina \citep{graham15} and Palomar Transient Factory (PTF) surveys \citep{charisi16}.

\begin{figure}
\includegraphics[width = 0.93\columnwidth]{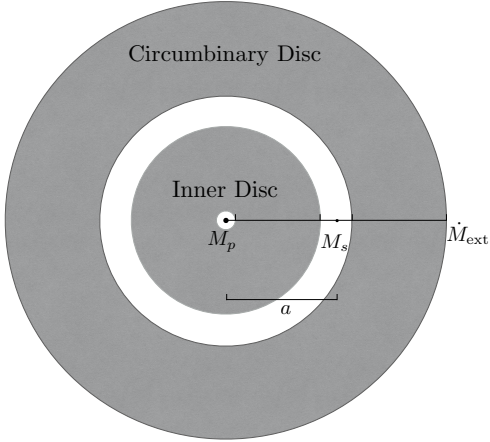}
\caption{Schematic representation of the binary--disc system for $q \ll 1$.
The whole system is surrounded by the circumbinary (outer) disc, while the primary black hole (mass $M_p$) has an individual (inner) disc extending out to the secondary. For simplicity, we neglect the accretion disc around the secondary black hole (mass $M_s$) and assume the BH and gas orbits are circular, co-planar, and pro-grade. The binary separation $a$ and the surface density $\Sigma(r)$ of the discs are coupled and evolve over time. The external boundary condition of the system is given by a fixed accretion rate $\dot{M}_{\rm ext}$.}
\label{f:scheme}
\end{figure}

The large computational cost of simulating the long-term evolution of these systems makes 3D or even 2D models prohibitive. Therefore, here we approach the problem with an 1D idealized configuration. We modeled the ``inner" disc around the primary and the ``outer" circumbinary disc with the standard hydrodynamics equations. We also include a simple prescription to mimic the binary tidal torque \citep{armitage02, tazzari15}, and allow for some gas to cross from the outer to the inner disc. For simplicity, we do not model the individual disc around the secondary and neglect the change in the mass ratio due to accretion (as discussed and justified in \S~\ref{sec:boundary}). A schematic representation of the system is shown in Fig. \ref{f:scheme}.

In this scenario many processes occur simultaneously. The tidal interaction removes angular momentum from the binary and adds it to the circumbinary disc, removing some gas from the vicinity. On the other hand, the inner disc transfer its angular momentum to the binary, slowing its migration \citep[e.g.][]{papaloizou84}. Viscosity in both discs dissipates kinetic energy and drags the gas inwards \citep{shakura73, frank02}, which added to the dissipation effect of the tidal torque heats up the gas, shaping its electromagnetic radiation \citep[e.g.][]{lodato09}. 

Important properties of the system depend on $q \leq 1$, the mass ratio of the binary. \citet{dorazio16} found that for $q \lesssim 0.05$ the tidal interaction digs an annular gap at the location of the secondary orbit, that the disc remains axisymmetric, and drives a constant accretion rate. On the other hand, for $q \gtrsim 0.05$, the gap becomes a central cavity around the binary, which furthermore becomes lopsided when $q \gtrsim 0.3$. Short-term 3D simulations of this process show however that streams of material get across the cavity and drive accretion episodically \citep{artymowicz94, cuadra09, roedig11, farris14, farris15, dunhill15, tang17}, although the efficiency of this feeding decreases as the discs become thinner \citep{ragusa16}. A similar result has been found in 2D and 3D simulations of $q \ll 1$ binaries, where material can cross the annular gap \citep{baruteau12, duffell14, duermann15, duermann17}, again depending on the disc thickness \citep{cerioli16}.

Several previous studies have focused on the long-term evolution of the binary--disc systems in the regime where $q \ll 1$, but they are either analytical, assuming steady state \citep{liu10, kocsis12b, rafikov16}, or numerical, but with a relatively small initial binary separation ($a \leq 10^4 ~{\rm R_S}$) and without inflow through the secondary orbit \citep{armitage02, lodato09, chang10, tazzari15}. In this work, we overcome these limitations by developing a self-consistent, time-dependent model that follows the long-term coupled evolution of the system starting at $10^5 ~{\rm R_S}$, including a prescription for material to cross the gap. 

This paper is organized as follows. In \S\ref{sec:phys} we present and explain the physical properties of the system we are considering, and present the equations for each process implemented in our numerical simulation. In \S\ref{sec:res} we show our main results on the evolution of a binary system, the surface density profile, the residence time, and the expected spectrum of the discs. In \S\ref{sec:pw} we compare our results with previous analytical and numerical works, and discuss the similarities and differences between our approach to the problem. Finally, in \S\ref{sec:con}, we present a summary of our main results and their implications.

Our most significant conclusion is that, at least for the mass ratios and inflow speeds included in our study, we find no real steady-state on the discs: the surface density distribution chosen as the initial condition influences the state of the discs during their whole subsequent history.

\section{System physics and model properties} \label{sec:phys}

In this section, we first explain the relevant physics in our study, then we discuss the simplifications needed to implement a 1D simulation, and finally give details of our numerical code.

We model the inner-- and circumbinary discs assuming their scale height $h$ is much smaller than the distance to the centre $r$, so the system behaves as a standard, two-dimensional thin disc. Furthermore, we consider the primary black hole to be at the center of mass of the system, ignore the individual disc around the secondary, and assume that the secondary and both accretion discs follow circular, co-planar and prograde paths around the primary black hole. These assumptions allow us to model the system in one dimension, with all disc properties functions of the radial coordinate $r$ only.

\subsection{Viscosity, thickness and energetics} \label{sec:visc}

Viscosity acts as an angular momentum transport mechanism, which produces a torque between contiguous rings of the disc. The standard $\alpha$-disc model \citep{shakura73} parametrizes the turbulent kinematic viscosity as a function of the sound speed $c_s$ and the thickness of the disc $h$. While $c_s$ depends on the total pressure, in this study we implement an alternative viscosity prescription called a $\beta$-disc model, which ensures thermal and viscous stability:
\begin{equation}
\nu = \alpha c_s h \beta, \label{eq:nu}.
\end{equation}
This allows us to put all the uncertainties in the constant $\alpha \leq 1$. Here, $\beta$ is the ratio between the gas pressure $p_{\rm gas}$ and total pressure $p_{\rm tot} = p_{\rm gas} + p_{\rm rad}$ (with $p_{\rm rad}$ the radiation pressure). For simplicity, we consider the pressures to be determined by the mid-plane temperature of the disc $T_c$: $p_{\rm rad} = 4 \sigma T_c^4 / (3 c)$ and $p_{\rm gas} = \rho k T_c / (\mu m_p)$, with $k$ the Boltzmann constant, $\sigma$ the Stefan-Boltzmann constant, and $\mu = 0.615$ the mean particle mass in units of the proton mass $m_p$ for a plasma of solar metallicity.

The sound speed in the disc is given by $c_s^2 = p_{\rm tot} / \rho$ with $\rho = \Sigma / (2 h)$ the volumetric density and $\Sigma$ the surface density of the discs. Assuming hydrostatic equilibrium in the vertical direction, we can express the sound speed as a function of thickness $h$ and angular velocity: $c_s = \Omega h$, with $\Omega = (GM(1+q)/r^3)^{1/2}$ the angular velocity of the material.

To close this system of equations, we relate the central temperature $T_c$ to the surface density $\Sigma$, assuming photons are transported to the disc surface by vertical diffusion: $F = 8 \sigma T_c^4 / (3 \Sigma \kappa)$ (where $\kappa$ is the opacity \added{due to electron scattering}), and that viscous and tidal heating are dissipated in the form of radiation \citep{lodato09, kocsis12a},
\begin{equation}
F = D_\nu + D_\Lambda = \frac{9}{8} \nu \Sigma \Omega^2 - \frac{1}{2} \Lambda \Sigma (\Omega - \Omega_s). \label{eq:F}
\end{equation}
The tidal term $D_\Lambda$, as explained in \citet{lodato09}, comes from angular momentum conservation and from the assumption that orbits remain circular. 

Using the above formulation leads to a system of equations, uniquely determining the disc properties at each radius and time:
\begin{equation} 
\begin{aligned}
T_c &= \Bigg[\dfrac{3 \kappa \Sigma^2 \big(9 \alpha c_s^2 \Omega \beta - 4 \Lambda(\Omega - \Omega_s)\big)}{64 \sigma}\Bigg]^{\frac{1}{4}},\\
\beta &= \left[1 + \frac{8 \sigma \mu m_p T_c^3 c_s}{3 c k \Sigma \Omega}\right]^{- 1},\\
c_s &= \frac{8 \sigma T_c^4}{3 c \Omega \Sigma (1 - \beta)}.
\end{aligned}\label{eq:solution}
\end{equation}
It can be shown that there is only one real and positive solution for the central temperature. Solving the equations allows us to obtain the sound speed $c_s$, the thickness $h$ of the disc, and finally the viscosity itself \citep{fontecilla17}. Having the central temperature at each radius and assuming that the discs emit as a multi-temperature blackbody \citep{frank02}, we can obtain the spectral energy distribution (SED) and the bolometric luminosity ($L_{\rm bol}$) of the system as a function of time.

\subsection{Surface Density evolution}

At least two mechanisms will make the surface density of the discs evolve over time: viscosity, already explained in the previous subsection, and the tidal torque produced by the changing gravitational potential of the rotating binary. The latter is a 2D effect, and cannot be implemented directly in a 1D simulation. To bypass this, we adopt a commonly used recipe that models its effect on the accretion discs. Following the literature \citep{armitage02}, 
\footnote{See \citet{dong11a, dong11b, petrovich12, rafikov16} for discussion about the tidal prescription.}
we define an orbit-averaged torque,
\begin{align}
\Lambda = \begin{cases}
- \dfrac{f}{2} q^2 \Omega^2 r^2 \bigg(\dfrac{r}{\Delta}\bigg)^4, & \mbox{if } r < a \\[12pt]
\enskip \dfrac{f}{2} q^2 \Omega^2 r^2 \bigg(\dfrac{a}{\Delta}\bigg)^4, & \mbox{if } r \geq a \\
\end{cases}\label{eq:smooth}
\end{align}
whose functional form depends on whether we are in the inner ($r < a$) or circumbinary ($r \geq a$) discs. Here \added{$f = 0.01$} is a dimensionless parameter that controls the strength of the torque, $\Delta = \max \{R_h,~h,~|r - a|\}$ is a smoothing term, and $R_h = a (q / 3)^{1 / 3}$ is the Hill radius of the secondary black hole. While this model was originally proposed for $q \ll 1$, it has been widely used for binaries up to $q = 0.3$ \citep{lodato09, chang10, kocsis12a, kocsis12b, tazzari15}.

While viscosity always transfers the angular momentum of the disc outwards, the tidal effect depends on the position. Inside the binary orbit ($r < a$) the tidal torque $\Lambda$ adds angular momentum to the binary, producing an inward acceleration on the elements of the disc. On the other hand, for the circumbinary disc ($r \geq a$), this effect will add angular momentum to the gas, preventing it to cross the binary orbit. Here, the balance between the tidal effect and the viscous mechanism will produce a low density region called cavity or gap, which depends on the mass ratio of the binary. 

As pointed out by \citet{tazzari15}, for $q \geq 0.1$, eq. (\ref{eq:smooth}) gives an unphysically large tidal contribution on the discs outside the Lindblad resonances, which will alter the migration velocity of the binary and the surface density of the discs. Following their work we added an exponential cutoff for the tidal torque outside this region,
\begin{align}
\Lambda = \begin{cases}
- \dfrac{f}{2} q^2 \Omega^2 r^2 \bigg(\dfrac{r}{\Delta}\bigg)^4 \exp\left[- \left(\dfrac{r - r_{\rm IMLR}}{w_{\rm IMLR}}\right)^2\right] & \mbox{if } r \leq r_{\rm IMLR}, \\[18pt]
\enskip \dfrac{f}{2}q^{2}\Omega^{2}r^{2}\bigg(\dfrac{a}{\Delta}\bigg)^4 \exp\left[-\left(\dfrac{r-r_{\rm OMLR}}{w_{\rm OMLR}}\right)^2\right] & \mbox{if } r\geq r_{\rm OMLR}.
\raisetag{38pt}
\end{cases}\label{eq:smooth2}
\end{align}
Here, $r_{\rm IMLR} = 0.63 a$ and $r_{\rm OMLR} = 1.59 a$ are the radius of the innermost and outermost Lindblad resonances, while $w_{\rm IMLR} = 370 h$ and $w_{\rm OMLR} = 75 h$ are the widths of the Gaussian smoothing. These values were used by \citet{tazzari15} to reproduce the gap sizes of a $q = 0.11$ binary from \citet{artymowicz94}.

Following \citet{frank02}, considering the mass continuity and angular momentum conservation equations in the discs, adding the viscous and tidal torques (eq. \ref{eq:smooth}), the surface density evolution can be written as:
\begin{equation}
\frac{\partial \Sigma}{\partial t} = - \frac{1}{r}\frac{\partial}{\partial r}\left( - 3 r^{1 / 2}\frac{\partial}{\partial r}\left(\nu \Sigma r^{1 / 2}\right) + 2 \frac{\Sigma \Lambda}{\Omega}\right), \label{eq:sigma}
\end{equation}
where the first term at the right hand side is the effect of viscosity and the second one is produced by the tidal interaction between the binary and the discs. 

\subsection{Binary separation evolution}

The separation between the SMBHs will evolve over time by two different mechanisms. The first one, mostly relevant at large binary separation, is the back-reaction between the binary and the disc due to the tidal interaction explained above. The second effect, which dominates at smaller separations, is produced by the GW emission. The following equation for the evolution of the binary separation $a$ incorporates both effects:
\begin{equation} 
\frac{da}{dt} = - \frac{4 \pi}{a M_s \Omega_s} \int_{r_{\rm in}}^{r_{\rm out}} \Lambda \Sigma r dr - \frac{8}{5} q (1 + q) c \left(\frac{{\rm R_S}}{a}\right)^3.
\label{eq:dadt}
\end{equation}
Here the first term of the right-hand side is the binary--disc interaction integrated over the whole disc, and the second term is the GW energy loss \citep{peters64}; $M_s = q M_p$ is the secondary's mass, $\Omega_s$ is its angular velocity, ${\rm R_S} = 2 G M_p / c^2$ is the Schwarzchild radius of the primary BH, $G$ the gravitational constant, and $c$ the speed of light.

The GW emission always tends to bring the SMBHs together. On the other hand, the presence of material inside the binary orbit will push the secondary outwards, while the material in the circumbinary disc will push it inwards. Therefore, the time until coalescence will be influenced by how material is distributed in the discs and how much of it can cross the cavity.

\subsection{Initial and boundary conditions} \label{sec:boundary}

After obtaining the equations for the surface density and binary separation evolution, we need a few more elements in our implementation to fully simulate the system. First, we need to specify the initial and spatial boundary conditions of the discs.

The initial surface density profile of the two discs was defined using a standard Shakura--Sunyaev accretion disc \citep{kocsis12a} with an accretion rate $\dot{M}_{\rm ini} = 4 \dot{M}_{\rm Edd}$ ($\dot{M}_{\rm Edd} = 4 \pi G M_p / (\kappa c \eta)$ is the Eddington accretion limit, with $\eta = 0.1$ the radiative efficiency). To include the effect of the secondary BH, we added an artificial cavity removing all the material within a factor of two of its initial position ($0.5 a_0 < r < 2 a_0$, with $a_0 = 10^5 ~{\rm R_S}$). We define a local disc mass as
\begin{equation}
M_{\rm loc} = \int_{a_0 - R_h}^{a_0 + R_h} 2 \pi r \Sigma_{\rm unp} dr, \label{eq:ldm}
\end{equation}
where $\Sigma_{\rm unp}$ is the expected surface density of the disc without the secondary. The ratio between the local disc mass and the secondary BH mass $M_s$ determines the type of migration and its timescale \citep[e.g.][]{haiman09}. For our fiducial case with $\alpha = 0.1$, $M_p = 10^7 ~M_\odot$, $q = 0.1$, and $\dot{M}_{\rm ini} = 1 ~{\rm M_\odot yr^{-1}}$, we obtain $M_{\rm loc} / M_s \approx 8$, so we would expect to be in the disc-dominated type II migration regime (but this is not necessarily the case, see discussion below).

In the outer radius of the circumbinary disc we implemented a constant external accretion rate $\dot{M}_{\rm ext}$ which feeds the system during the simulation. For consistency, we use the same accretion rate for the initial condition, $\dot{M}_{\rm ext} = \dot{M}_{\rm ini}$.

For the inner radius of the individual disc around the primary BH we used a zero torque boundary condition, which allows material to be freely accreted onto the SMBH.

The outer boundary of the inner disc and the inner boundary of the circumbinary disc are naturally added in the model by the tidal effect implemented with eq. \ref{eq:smooth2}. This tidal torque will open a cavity around the secondary black hole, pushing the material in the discs away from its orbit.

For simplicity, we did not evolve the black hole masses in our model. However, we checked a posteriori if this assumption was justified. To do this we assumed (i) the initial disc inside the secondary orbit is accreted by the primary, and, (ii) all the material that crosses the cavity is accreted by one of the BH. At the end of the simulations we obtained, on average, a variation $\Delta q / q \approx 25 \%$ and $\Delta M_p / M_p \approx 34 \%$.

Many previous works have shown that the gap opened by the secondary is not an impenetrable wall: gas can cross the binary orbit and enter to the inner disc in the form of dense streams of material. This inflow is crucial because it feeds the inner disc, contributing to the hot part in the SED, determining how much gas will be left just before the coalescence and, indirectly, controlling the migration timescale. Since this process is (at least) a 2D effect, we cannot directly compute it, and we need to use a recipe to include it in our code. Some possibilities could be using a constant accretion \citep{kocsis12a, kocsis12b}, a fraction of the external accretion \citep{rafikov16}, or a fraction of the steady-state accretion from a standard single BH Shakura--Sunyaev disc. Using any of these alternative implies the inflow is tied directly to the viscous evolution of the circumbinary disc. However, the flow across the gap or cavity is produced by the tidal (gravitational) torques, and so it should operate on the orbital time of the binary, and not on the viscous time of the gas near the gap/cavity wall. Recent hydrodynamical simulations confirm this and suggest that this accretion process should be mostly dependent on the dynamical properties of the binary rather than on the local physics of the disc \citep{farris15, dorazio16}. Motivated by this, in our work we consider a different recipe, which allows us to control the rate at which material crosses the cavity, but ties this rate to the local orbital time:
\begin{equation} 
\dot{M}_{\rm cross} = \gamma \Omega_s r^2 \Sigma.
\label{eq:mdotc}
\end{equation}
Here $\gamma$ is a free parameter that determines the fraction of the nearby material that crosses the secondary radius per binary orbit. As an example, $\gamma = 10^{- 4}$ implies that, ignoring viscosity effects, it would take $\sim 10^4$ orbits to totally deplete the nearby circumbinary disc. To prevent large fluctuations in the accretion rate, eq. (\ref{eq:mdotc}) is implemented using the average values of the radius ($\bar{r}$) and the surface density $\left(\bar{\Sigma}\right)$ in the region of the circumbinary disc where the tidal torque dominates. Based on the 3D simulations of \citet{cuadra09} the range used was $a < r < 3 a$. From these radii we extract, once per binary orbit, a gas mass given by the product between the accretion rate and the orbital time: $\dot{M}_{\rm cross} \Delta t_{\rm orb}$. In general, some of this mass will cross the gap (e.g. on horse-shoe orbits for $q\ll 1$), reach the outer edge of the inner disc, and eventually accrete onto the primary BH. The remaining mass is captured and accreted onto the secondary (in our model, since we do not include the circum-secondary disc, we remove this material from the simulation). We adopt the ratio between these accretion rates from \citet{farris14}, who computed them explicitly in 2D hydrodynamical simulations over a range of $q$. Specifically, based on their Figure 7, for our simulations with $q = (0.1,~0.3,~0.01)$, we assume that $(6 \%,~40 \%,~85 \%)$ of the mass that leaves the circumbinary disc reach the inner disc. For simplicity, we add this mass at a fixed radius, i.e. as a delta function at $r/a = (0.4,~0.2,~0.6)$, based on the tidal truncation radii obtained by \citet{artymowicz94}. Since we do not modify the masses of the BHs, these values are fixed during the whole simulation. 

\added{When we add the gap-crossing material to the inner disc, we endow it with the local Keplerian velocity at the new location. 
As a result 
our simple inflow implementation does not conserve angular momentum (even if we were to add all the mass to the inner disc).
Without following the non-axisymmetric gas flows in 2D or 3D, we do not know where this numerically lost angular momentum would end up.  So in order to estimate if this could affect our results, we consider two extreme scenarios: the lost angular momentum could have been (i) added to the secondary BH, or (ii) deposited back in the circumbinary disc. In either case, we found that for $\gamma \leq 10^{-5}$ the effects are sub-dominant compared to either (i) the binary's orbital evolution in the current simulation, or (ii) the viscous angular momentum flux at the inner edge of the circumbinary disc.}

\subsection{Model parameters and numerical implementation}

\begin{table*}
 \caption{Simulation properties. From left to right the columns are: a reference simulation number, the logarithm of the primary mass in solar masses, the binary mass ratio ($q$), our parameter to control the inflow through the gap ($\gamma$), the Shakura--Sunyaev viscosity parameter ($\alpha$), the initial and external accretion rate in units of the Eddington limit, the grid resolution, the time resolution, and the differences between the current simulation and our fiducial case (simulation $N^{\rm o} 1$). See text for a more detailed explanation.
 }
	\label{tab:params}
  \begin{tabular*}{0.96\textwidth}{lcccccccc} 
		\hline
    \\[-2ex]
		$N^{\rm o}$ & log$\left(\dfrac{M_p}{M_\odot}\right)$ & $q$ & $\gamma$ & $\alpha$ & $\dfrac{\dot{M}}{\dot{M}_{\rm Edd}}$ & $n$ & $t^*$ & Differences\\
    \\[-2ex]
		\hline
    \\[-2ex]
    $~~1$ & $7$ & $0.1$ & $10^{- 4}$ & $0.1$ & $4$ & $500$ & $10^6$ & - \\ 
    \hline
    \\[-2ex]
    $~~2 - 7$ & - & - & $10^{- 5},~10^{- 6},~10^{- 7},~10^{- 8},~10^{- 9},~0.0$ & - & - & - & - & $\gamma$ \\ 
    $~~8 - 21$ & - & $0.3,~0.01$ & $10^{- 4},~10^{- 5},~10^{- 6},~10^{- 7},~10^{- 8},~10^{- 9},~0.0$ & - & - & - & - & $\gamma$ and $q$ \\ 	 
    $22 - 23$& - & - & - & $0.01,~1$ & - & - & - & $\alpha$ \\ 
    $24 - 25$ & - & - & - & - & $0.4,~40$ & - & - & $\dot{M}$ \\ 
    $26 - 28$ & $6,~8,~9$ & - & - & - & - & - & - & $M_p$ \\
    $29$ & - & - & - & - & - & -& - & without inner disc \\ 
    $30$ & - & - & - & - & - & -& - & without both discs \\ 
    $31$ & - & - & - & - & - & $10^3$ & $10^5$ & increase resolution \\
    $32 - 36$ & - & $0.1$, $0.01$ & $10^{- 5},~0.0$ & - & $0.1$ & - & - & \citeauthor{kocsis12b} parameters \\
    $37 - 38$ & - & $0.01$ & $10^{- 5}$, $0.0$ & $0.3$ & $0.1$ & - & - & \citeauthor{haiman09} parameters \\
    $39$ & $8$ & - & $0.0$ & $0.1$ & $1$ & - & - & \citeauthor{lodato09} parameters \\
    $40$ & - & - & - & - & - & - & - & empty discs, no secondary \\
    \hline     
	\end{tabular*}
\end{table*}

In Table \ref{tab:params}, we list the relevant properties of our simulation runs. The first row refers to our fiducial case. The rest of the table shows how each model deviates from this fiducial case. The first 21 simulations are the core of our work, and are the combinations between $q = \{0.1,~0.3,~0.01\}$ and $\gamma = \{10^{- 4},~10^{- 5},~10^{- 6},~10^{- 7},~10^{- 8},~10^{- 9},~0.0\}$. In the next set of simulations, we changed the $\alpha$ parameter ($22-23$), the external accretion rate ($24-25$) and the primary BH mass ($26-28$). We also started the simulation without either the inner or both discs ($29-30$) to explore the effects of the inflow from the outer boundary and of the initial conditions. We run a high-resolution simulation ($31$) to test the numerical convergence of our results. We have run simulations for the comparisons made in \S \ref{sec:pw} with previous work, with parameters following those in \citet{kocsis12b} ($32-36$), \citet{haiman09} ($37-38$) and \citet{lodato09} ($39$). Finally, as a basic test of our code, we ran an additional simulation without a secondary BH, with a constant $\dot{M}_{\rm ext}$, and initially empty discs ($40$). We recovered satisfactorily the Shakura--Sunyaev standard solution for the disc for the given accretion rate.

To convert into code units we define the following quantities: $R_0 = 10^2 ~{\rm R_S}$, $T_0 = 10^3 ~{\rm K}$, $\Sigma_0 = 10^7 ~{\rm g / cm^2}$, $t_0 = t^* 2 \pi / \Omega(R_0)$ with $t^* = 10^{6}$ a scale factor for the time coordinate, and $m_0 = \Sigma_0 R_0^2 \sim 4 \times 10^2 M_\odot$ for a primary BH mass of $10^7 M_\odot$. Then, we divide every physical constant (or variable) used in the simulation with a combination of the previous definitions to obtain dimensionless quantities. Using the dimensionless versions of the equations for the processes shown in the previous section, we compute the state of the discs at each snapshot. To solve the partial differential equation of the surface density evolution we used the backward Euler method, an implicit numerical integration procedure \citep{hein06}. This method requires that we solve a tridiagonal matrix, which we did using the Thomas algorithm (a special case of Gaussian elimination). We used a front-end code in \textit{python} to work with the arrays, to set the properties of the simulations, and for visualization. For the heavy calculations, such as the matrix definition and solution, we developed a library in \textit{C}. We connected the two types of codes using \textit{ctypes}.

The surface density of the discs at each snapshot is the key variable in the models: with it we can compute the central temperature $T_c$, the pressure ratio $\beta$ and the sound speed $c_s$ (Eq. \ref{eq:solution}). We then obtain the thickness $h$ of the discs and the viscous and tidal interaction (Eq. \ref{eq:sigma}), which in turn sets how fast the separation between the SMBHs shrinks (Eq. \ref{eq:dadt}). Once we obtain the updated values of these variables, we evolve the system by one time step, obtaining a new surface density distribution.

We used a linearly spaced grid for the temporal coordinate and a logarithmic grid for the radial coordinate. We defined a numerical resolution $n$, which translate into a radial step-size $\Delta r / R_0 = r / R_0 (1 - 10^{1 / n})$ and a temporal step-size $\Delta t / t_0 = 1 / n^2$. We have run simulations with different values of $n$, and chose $n = 500$, which we have found to provide a good compromise between precision, convergence and computing time for our study.

Since we cannot directly demonstrate that the logarithmic grid fulfills the conditions to make the truncation error go to zero in the Taylor series \citep{hein06}, for our fiducial case we also ran a simulation with a linear radial grid, and found no significant differences.

By definition, throughout the cavity, the surface density should go to zero. This is not always the case in the simulations, due to truncation errors that accumulate over time. To address this numerical noise, we added two control conditions. First, if $\Sigma / \Sigma_0 \leq \Sigma_{\rm err} = 10^{- 20}$ in some radial cell, we set the surface density of that cell to zero. We expect that, at a given snapshot, when a disc "ends" (i.e, the surface density drops to zero at some point), it does not increase again on the same side of the secondary BH orbit. Our second condition is therefore to find the first cell where $\Sigma / \Sigma_0 = 0$ in each disc, and, for every cell from that point up to the binary separation, we set their surface density to zero. We tested these two control conditions to see if they change the behaviour of the discs and found that, aside from the elimination of some artifacts in the cavity itself, the final results are almost the same with and without them.

\section{Simulation results} \label{sec:res}

In this section, we present the main results of our work, including the surface density evolution, the migration timescale, the residual mass left in the inner disc, and the electromagnetic emission of the system at different times.

\subsection{Evolution of the disc surface density ($q = 0.1$)}

\begin{figure*} 
\centerline{\includegraphics[width = \textwidth]{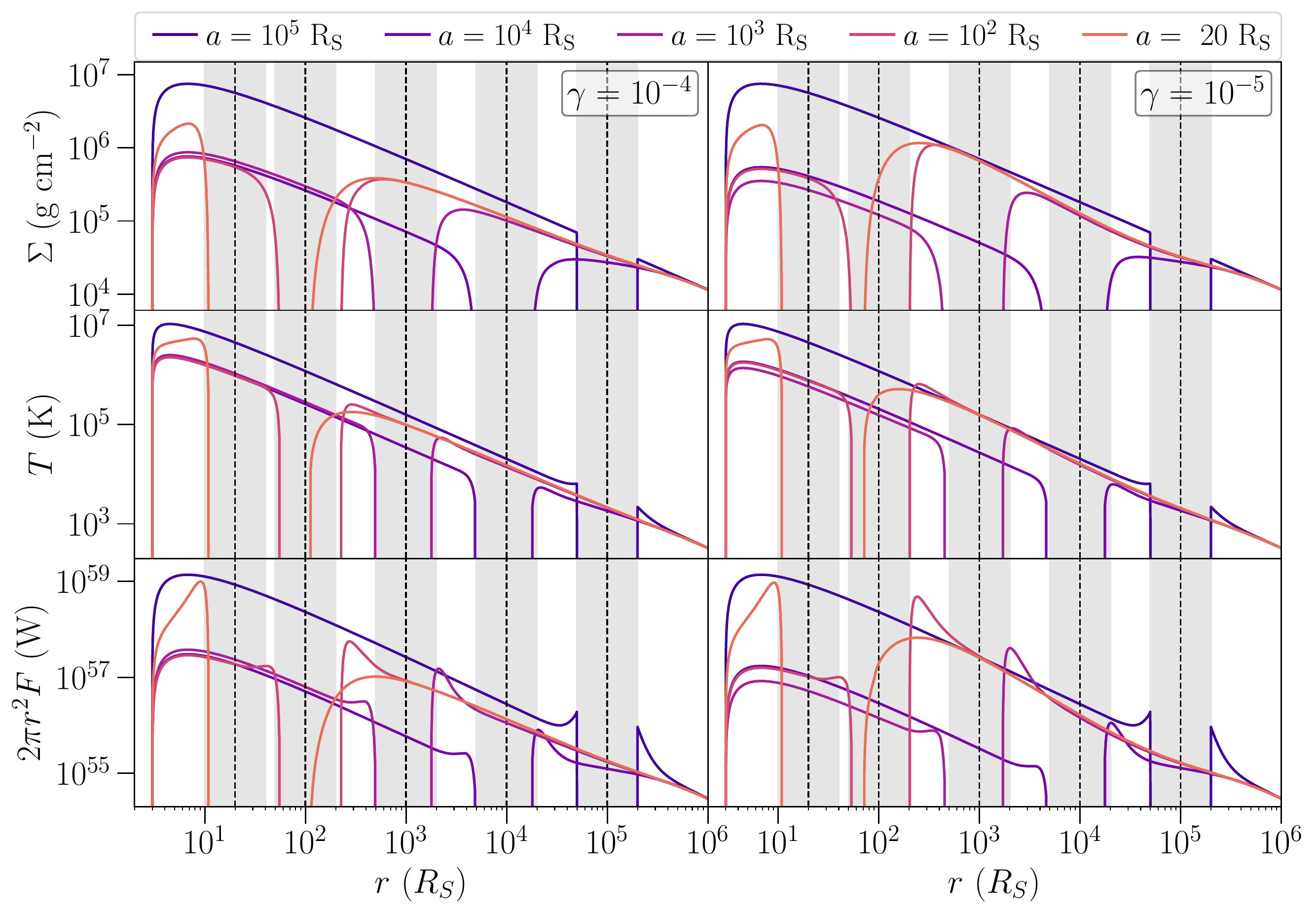}}
\caption{Results for a SMBHB of mass ratio $q = 0.1$ with $\gamma= 10^{- 4}$ (left, at: $13$ Myr, $4.7$ Myr, $830$ Kyr, $440$ yr and $0.7$ yr before merger) and $\gamma = 10^{- 5}$ (right, at: $9.6$ Myr, $2.5$ Myr, $340$ Kyr, $440$ yr and $0.7$ yr before merger). The black dashed vertical lines mark the position of the binary at a given separation and the gray regions represent the analytical size of the cavity ($0.5 a$ to $2 a$). The top panel shows the surface density $\Sigma$, the middle panel the midplane temperature $T_c$, and the bottom panel the surface brightness $2 \pi r^2 F$. While the $\gamma = 10^{- 4}$ case approaches a steady-state solution without any pile-up, in the case with $\gamma = 10^{- 5}$ the inner disc becomes depleted and then refills, while the circumbinary disc suffers a modest pile-up.}
\label{f:data}
\end{figure*}

In this subsection, we show the main results for the runs with the fixed fiducial mass ratio $q = 0.1$, but for different inflow rates controlled by the parameter $\gamma$. Figure \ref{f:data} shows the radial profiles of different relevant quantities for some of these models. The left panels correspond to the case with $\gamma = 10^{- 4}$, while the right panels correspond to $\gamma = 10^{- 5}$. Lines of different colours show the state of the system at different times, labeled by the corresponding binary separation.

The top row of Figure \ref{f:data} shows the evolution of the surface density $\Sigma$. The initial condition can be recognized by the sharp cutoff at the cavity edges and the high surface density in the inner disc (set by a constant accretion rate, \S~\ref{sec:boundary}). Once the binary separation has shrank to $a = 10^4 \ {\rm R_S}$, the slope of the surface density in the inner boundary of the circumbinary disc has decreased, since the migration timescale of the binary is shorter than the inflow timescale of the material ($t_{\rm flow} = r / |v_r|$) at $r \geq 4 \times 10^4 \ {\rm R_S}$, so it cannot refill the inner region of the circumbinary disc nor accumulate there.

At $a = 10^3 \ {\rm R_S}$, the decrease in the migration velocity allows the outer disc to recover its initial shape. This slow down in the shrinkage velocity is likely produced by the change in the available mass in the interacting region given by the Lindblad resonances (proportional to the binary separation $a$; \citealt{tazzari15}). On the other hand, the inner disc has not changed much, except that its outer boundary is depleted by the effect of the tidal torque.

Around $a = 10^2 \ {\rm R_S}$ we expect the so-called ``first decoupling'' to happen: at this point the inflow timescale of the circumbinary disc becomes longer than the binary migration timescale due to GW emission. The material of the outer disc will be no longer pushed away by the tidal torque and will evolve only due to viscosity. In this simulation, the inner disc still maintains the same surface density than the previous snapshots, showing that, for this $\gamma$, we find a configuration approaching a steady state.

During the GW emission phase the secondary will migrate inward faster than the viscous time of the inner disc, pushing the gas inward and squeezing it. The final curve shows the moment when the binary separation is $a = 20 \ {\rm R_S}$, where a so-called ``second decoupling'' should occur due to the thickening of the disc, which invalidates the thin-disc assumptions \citep{fontecilla17}. The figures shows that the gradient of the surface density in the inner region of the circumbinary disc becomes shallower, as it is not affected by the tidal barrier. The inner disc, on the other hand, becomes coupled with the secondary black hole at some point between the snapshots, and is squeezed, increasing its surface density by more than a factor of two.

In our simulations with this value of $\gamma$, we find that the slope of the circumbinary disc never becomes steeper than the single-BH case \citep[a process commonly known as pile-up, e.g.,][]{kocsis12a}. We attribute this to the fact that this $\gamma$ is large enough to constantly make the removed mass a considerable fraction of the available mass in the interaction region at the moment of extraction, preventing a pile-up. 

The right panel in the same row shows a simulation with the same mass ratio, but with $\gamma = 10^{- 5}$. The reduction in the rate at which material crosses the cavity produces a significant change in the discs: the inner disc becomes slightly depleted at some point after $a = 10^4 \ {\rm R_S}$, and the circumbinary disc can pile up gas, increasing its surface density and speeding up the migration. The inner disc refills again before the binary reaches the separation of $a = 10^2 \ {\rm R_S}$, showing that the depletion and pile-up do not produce a large difference in the surface density distribution of the inner disc at the end of the simulation, the residual mass (see Fig. \ref{f:myt} below) or the hot part of the SED.

The differences between $\gamma = 10^{- 4}$ and $\gamma = 10^{- 5}$ can be understood by defining an inflow timescale across the gap, $t_{\rm cross} \equiv 2 \pi r^2 \Sigma / \gamma \Omega_s \bar{r}^2 \bar{\Sigma}$. For $\gamma = 10^{- 4}$, we find that, in the interaction region in the circumbinary disc, $t_{\rm cross}$ is smaller than the background inflow timescale $t_{\rm flow}$ of the material in the circumbinary disk, and no pile-up can happen, whereas for $\gamma = 10^{- 5}$, $t_{\rm cross}$ becomes slightly longer than $t_{\rm flow}$, suggesting that a pile-up must occur.

The exact value of $\gamma$ for which our inflow model results in steady-state accretion can be estimated by setting $t_{\rm cross} = t_{\rm vis} = 2 r^2 / (3 \nu)$. Considering, for simplicity, an $\alpha$ prescription for the viscosity, 
\begin{equation}
\gamma = 3 \pi \alpha \left(\frac{h}{r}\right)^2 \frac{\Omega}{\Omega_s} \label{eq:gammass},
\end{equation}
for $\alpha = 0.1$, $h / r = 10^{- 2}$ and $r = 2 a$, we obtain $\gamma \sim 3 \times 10^{- 5}$. We obtain a similar value when we calculate this $\gamma$ directly from our fiducial simulation. At $r\sim 2a$, $\beta \ll 1$ and eq. \ref{eq:gammass} explains why the surface density profile presents no pile-up and looks similar to a steady-state solution.

\subsection{Evolution of the disc temperature and spectrum ($q = 0.1$)}

The temperature and surface brightness for the same binary separations are shown in the second and third row of Fig.~\ref{f:data}. Their behaviour is similar to the one of the surface density, as expected.
\added{We note that in the outer parts of the circumbinary disc, the temperature falls below $10^4$ K, where our simple opacity implementation due to electron scattering will not be sufficient to properly model the hydrogen ionization instability that should occur.}

\begin{figure}
\includegraphics[width = 0.93\columnwidth]{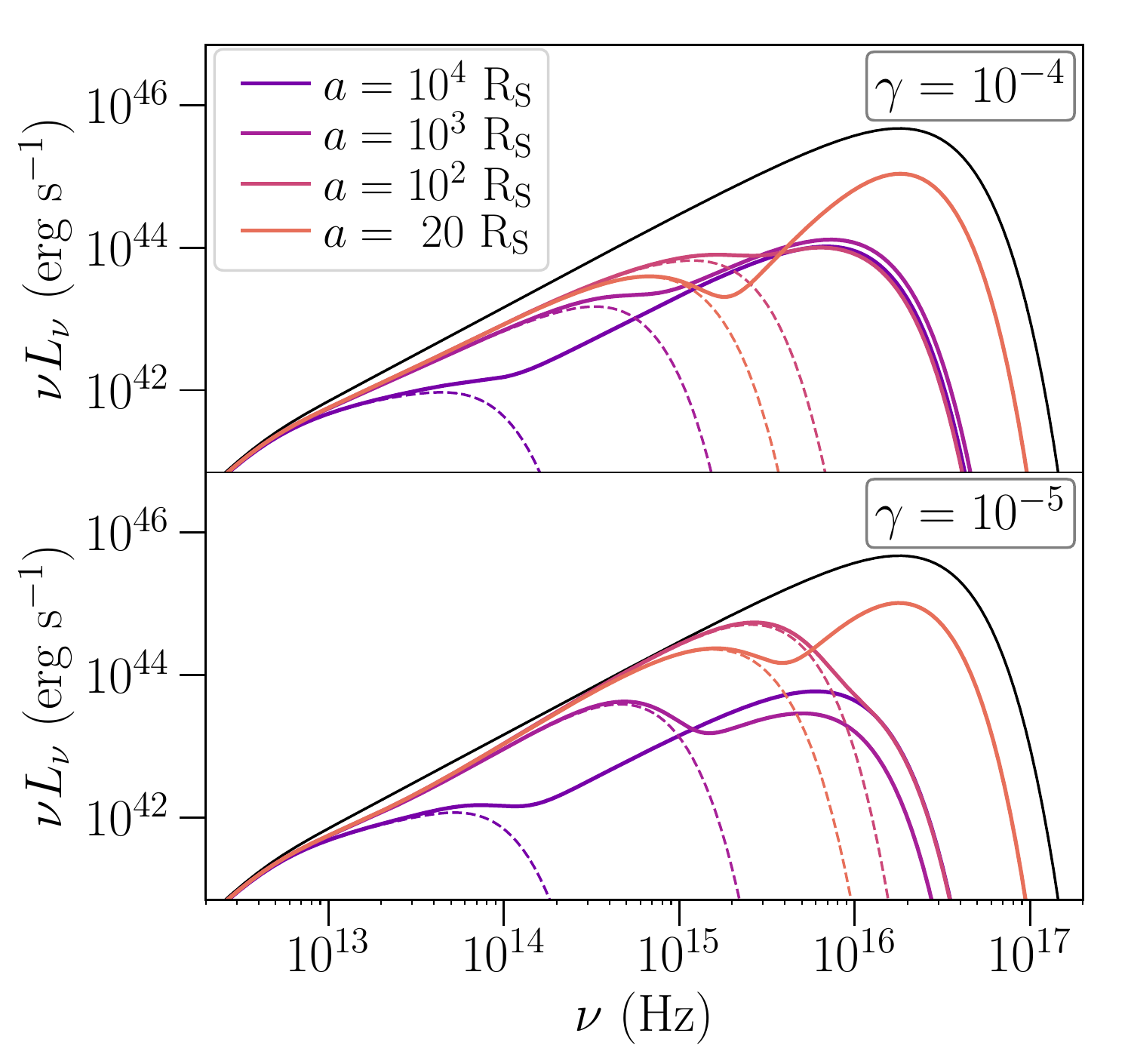}
\caption{Spectral energy distributions (SED) from the inner and outer discs, obtained assuming they are optically thick and emit as a multi-temperature black body. The top panel shows the results for $q = 0.1$ and $\gamma = 10^{- 4}$ while the bottom panel shows $\gamma = 10^{- 5}$. The thick curves correspond to combined total SED, with the black curve the case of a single-BH with $\dot{M} = \dot{M}_{\rm ext}$ for reference. The thin dashed curves indicate the SED from the circumbinary disc alone. Depending on the rate at which material crosses the secondary's orbit, the peak of the SED from each disc can dominate at different stages of the binary evolution.}
\label{f:sed}
\end{figure}

Using the temperature profiles, we compute the electromagnetic (EM) emission for the two cases in Fig.~\ref{f:data}. The top panel of Fig.~\ref{f:sed} shows the SED for $\gamma = 10^{- 4}$ at the same binary separations as in Fig.~\ref{f:data}. For comparison, a single-BH SED is shown in black. The formation of a cavity due to the presence of the binary produces a conspicous feature -- a down-ward depression, or ``notch'', in the spectrum \citep[e.g.,][]{gultekin12}. The (hotter) inner disc dominates the SED to the right of the notch (i.e higher frequencies), while the (colder) circumbinary disc dominates to the left. As the system evolves, the outer disc can reach closer to the primary BH, heating up and showing an enhancement at shorter wavelengths. The notch in the SED caused by the cavity also moves to the right while the inner disc almost preserves its shape, until $a = 10^2 \ {\rm R_S}$, where the squeezing starts and the EM emission peaks at higher frequencies \citep{farris15, tang18}.

The case with $\gamma = 10^{- 5}$ is shown in the bottom panel of the same figure. The behaviour for this simulation is similar, but the pile-up of the material at the inner boundary of the circumbinary disc produces an enhanced contribution \citep{lodato09, kocsis12b}. For this $\gamma$, the dominance of the inner disc is not permanent: the circumbinary disc surpasses its luminosity after $a = 10^3 \ {\rm R_S}$. However, the inner disc dominates again after the first decoupling.

The frequency at which the notch appears in the SED is also different in the two cases, because removing material from the circumbinary discs also changes the effective position of the cavity. In the case with smaller $\gamma$, this allows more material to be hotter and closer to the secondary BH.

\begin{figure}
\includegraphics[width = 0.93\columnwidth]{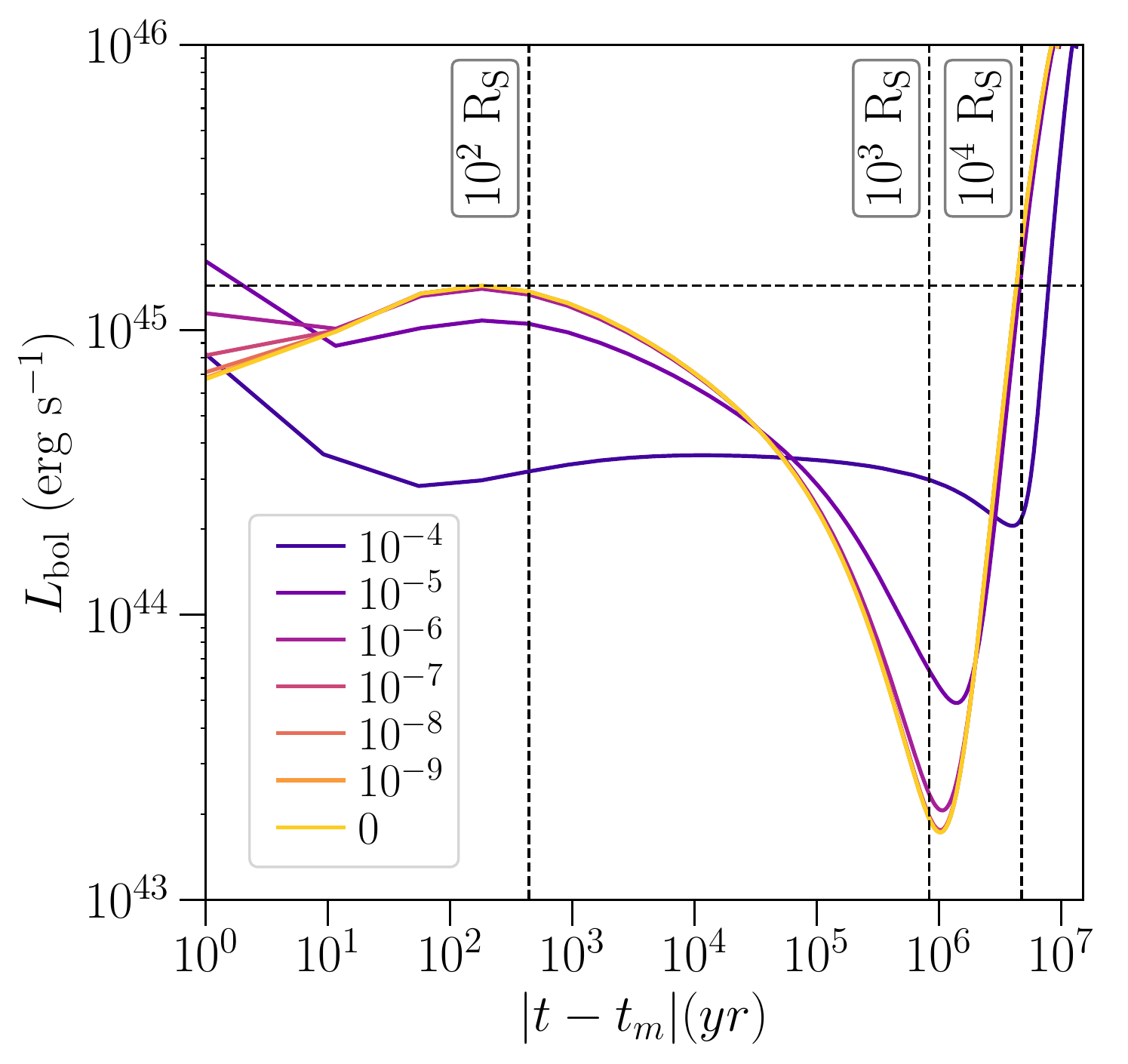}
\caption{Bolometric luminosity ($L_{\rm bol}$) for the simulations with $q = 0.1$ and each value of $\gamma$ as a function of time (running from right to left). The vertical dashed lines mark different binary separations for the case with $\gamma = 10^{- 4}$, and the horizontal dashed line shows the Eddington luminosity for this system. For the cases when material crosses the cavity rapidly ($\gamma \geq 10^{- 5}$), there is relatively more gas near the primary close to the merger, and the peak at the second decoupling ($\sim$1 year prior to merger) is higher than at the first decoupling ($\sim$100 days prior to merger). On the other hand, when the secondary ``dam'' is more efficient ($\gamma \leq 10^{- 6}$), the circumbinary disc pile-up, peaking at the time of the first decoupling, produces the brightest luminosity during the system's evolution.}
\label{f:bol}
\end{figure}

Summing over all frequencies, we obtain the bolometric luminosity $L_{\rm bol}$. The results for each of the $q = 0.1$ simulations are shown in Fig.~\ref{f:bol}, as a function of time before coalescence. The vertical dashed line indicates the binary separation for the case with $\gamma = 10^{- 4}$, while the horizontal line marks the Eddington luminosity for this system.

Roughly half of the time in the binary evolution is spent going from the initial separation at $a = 10^5 \ {\rm R_S}$ to $a = 10^4 \ {\rm R_S}$. At the end of the simulation, all systems become brighter compared to their luminosity at $a \sim 10^4 \ {\rm R_S}$, due to the squeezing of the inner disc. Since, as pointed out by \citet{lodato09}, we include the tidal heating in both discs, the inner region of the circumbinary disc can dominate the bolometric emission showing a peak around the time the first decoupling happens.

There are some important, visible differences between the models with different $\gamma$: the simulation with $\gamma = 10^{- 4}$ shows a more constant luminosity, since the material crosses the cavity sufficiently rapidly to prevent a pile-up in the circumbinary disc. For $\gamma = 10^{- 5}$, a pile-up is formed, but material still crosses the cavity rapidly enough to make the emission at the squeezing end-phase ($\sim$1 year prior to merger) brighter than the one at the first decoupling. Finally, for $\gamma \leq 10^6$ this does not happen, and the system is brightest during the first decoupling, when the pile-up is most pronounced (seen at $\sim$100 days prior to merger in Fig.~\ref{f:bol}).

The enhancement in the luminosity after its minimum is due to the refiling of the inner disc and/or the pile up of the circumbinary disc. After the first decoupling the luminosity decreases again, because the outer disc is no longer influenced by the binary's tidal torques and heating. Reaching the end of the simulation, the secondary BH squeezes the inner disc, increasing the overall luminosity of the system, but, for $\gamma \leq 10^6$ there is a relatively small amount of gas left near the primary to be heated, and the luminosity peaks around the first decoupling.

\subsection{Binary inspiral timescales}

We next compare the main results among the first 21 simulations, to show how the variation in $q$ and $\gamma$ affects the time the binary spends at each radius and how much material is left in the inner disc when the first decoupling occurs.

\begin{figure}
\includegraphics[width = 0.93\columnwidth]{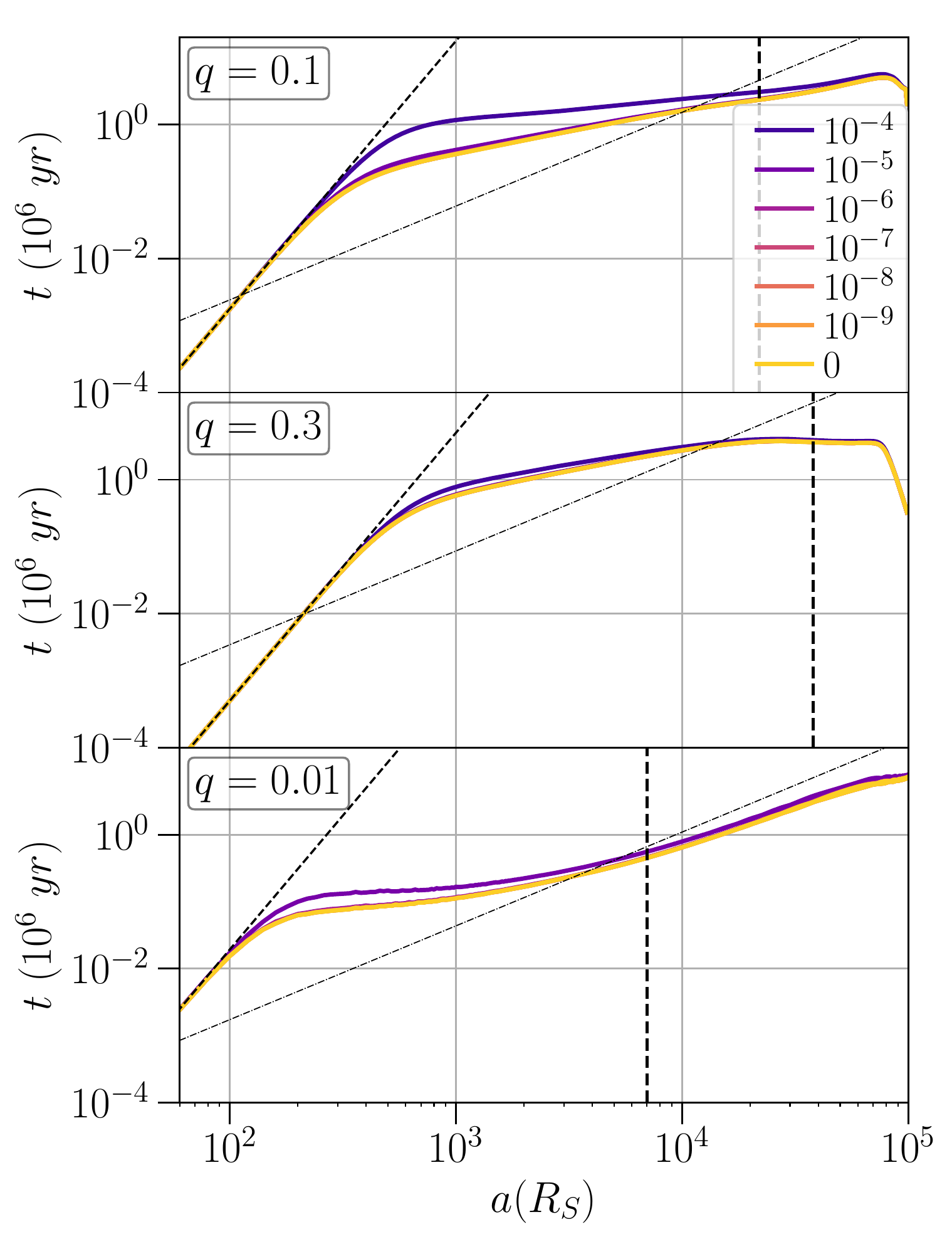}
\caption{Residence time ($t_r \equiv a / (da / dt)$) spent by the binary at each distance $a$ from the primary, shown for each of the $\gamma$ values used in this work. Top panel: $q = 0.1$, middle panel: $q = 0.3$ and bottom panel: $q = 0.01$. The behaviour at $a \sim 10^5 \ {\rm R_S}$ is produced by the redistribution of the material in the boundaries of the cavity from the sharp edges of the initial condition. At the end, when $a \sim 5 \times 10^2 \ {\rm R_S}$, GW emission starts to dominate the torques from the gas disc (the diagonal line shows a pure GW residence time). The light diagonal line is the viscous timescale of the circumbinary disc while the vertical thick dashed line marks the separation where the local disc mass becomes equal to the secondary's mass.}
\label{f:rest}
\end{figure}

Fig. \ref{f:rest} shows the residence time $t_{\rm r} \equiv a / (da / dt)$, as a function of the binary separation $a$. The thin dashed black line, which every simulation approaches at the end, displays the residence time assuming only GWs were present, while the thin dashed gray line shows an analytical estimate of the viscous time in the inner regions of the circumbinary disc, at the same binary separations. The thick vertical dashed line, on the other hand, indicates the position where the local mass of the disc, defined by eq. \ref{eq:ldm}, becomes smaller than the secondary BH mass.

At the beginning of the simulations, the effect of the sharp initial condition is clear: the system needs time to create the right cavity shape before it starts to evolve more smoothly. Following this adjustment, we can distinguish three migration regimes: At the right of the vertical dashed line, the binary should be following a disc-dominated type II migration. When the local disc and secondary masses become similar, the system should change to a slower secondary-dominated type II migration \citep{haiman09}. This is especially visible in the bottom panel, as a flattening of the slope at this transition. Around the same binary separation, the viscous timescale of the circumbinary disc becomes smaller than the residence time, allowing the material to pile up over time. After the binary separation becomes smaller than $a \sim 500 \ {\rm R_S}$, we see another clear steepening in the slope of each curve, showing the beginning of the GW emission regime.

The top panel shows the simulations with $q = 0.1$ for every value of $\gamma$. Except for the case with $\gamma = 10^{- 4}$, all the other cases have a similar behaviour. As we argued above, values higher than $\gamma = 10^{- 5}$ have accretion timescales shorter that the inflow timescale. 

\begin{table}
\centering
	\caption{Disc properties related with the type of migration. From left to right, the initial disc mass of the inner disc, the initial local mass around the secondary and the binary separation when the local mass becomes equal to the secondary mass}
	\label{tab:dprop}
  \begin{tabular*}{0.7\columnwidth}{lccc} 
     $q$ & $M_{\rm int} ~(M_s) $ & $M_{\rm loc} ~(M_s)$ & $a ~(\rm{R_S})$ \\ 
    \\[-2ex]
		\hline
    \\[-2ex]
    $0.1$ & $~~3.38$ & $~~8.09$ & $2.25 \times 10^4$\\
    $0.3$ & $~~1.13$ & $~~3.87$ & $3.81 \times 10^4$\\
    $0.01$ & $33.82$ & $37.30$ & $7.40 \times 10^3$\\
    \hline     
	\end{tabular*}
\end{table}

The bigger deviation happens during the secondary-driven evolution, where the contribution of the circumbinary disc becomes less and less important due to the shrinkage of the interaction zone.

When the binary has more similar masses ($q = 0.3$, middle panel) all simulations follow the same trend: more time is needed at the beginning to reach a more physical surface density distribution from the initial condition, while at the end the GW emission becomes dominant faster. We acknowledge that this mass ratio is at the limit of our capabilities with a 1D code, since considering the primary BH to be fixed at the origin, or using the tidal torque implementation, are no longer good approximations. 

Finally, the bottom panel shows the case with a smaller secondary BH ($q = 0.01$). For this mass ratio, we do not show results for $\gamma = 10^{- 4}$, because in that case the material that is partially added to the inner disc (and removed from the circumbinary disc) changes the direction of the binary migration, increasing their separation instead of shrinking it (see Eq. \ref{eq:dadt}). We suspect this result is an artefact of our initial condition and inflow implementation: the sharp initial gradients cause an inflow much more rapid than expected from the viscous timescale. We dismiss this behaviour as numerical and not physically relevant.

Table \ref{tab:dprop} summarizes the initial disc mass in the form of the mass in the inner disc ($M_{\rm int}$) and as the local mass $(M_{\rm loc})$), along with the separation when the latter becomes equal to the secondary mass. For each mass ratio, we compared the slope of the residence time obtained here to the analytical estimations from \citet{haiman09}. We found that, except for $q = 0.01$, our slopes are flatter than the one expected for disc-dominated migration. The results suggest that the condition for this type of migration is that the disc mass needs be higher than the secondary mass by at least an order of magnitude.

We define a ``residual mass" as the mass in the inner disc at the time when the binary separation reaches $ a = 100 {\rm R_S}$, which is when we expect the first decoupling and the inflow through the cavity stops (or at least slows down). This is similar to the ``fossil disc mass'' defined by \citet{chang10} and \citet{tazzari15}. 

\begin{figure}
\includegraphics[width = 0.93\columnwidth]{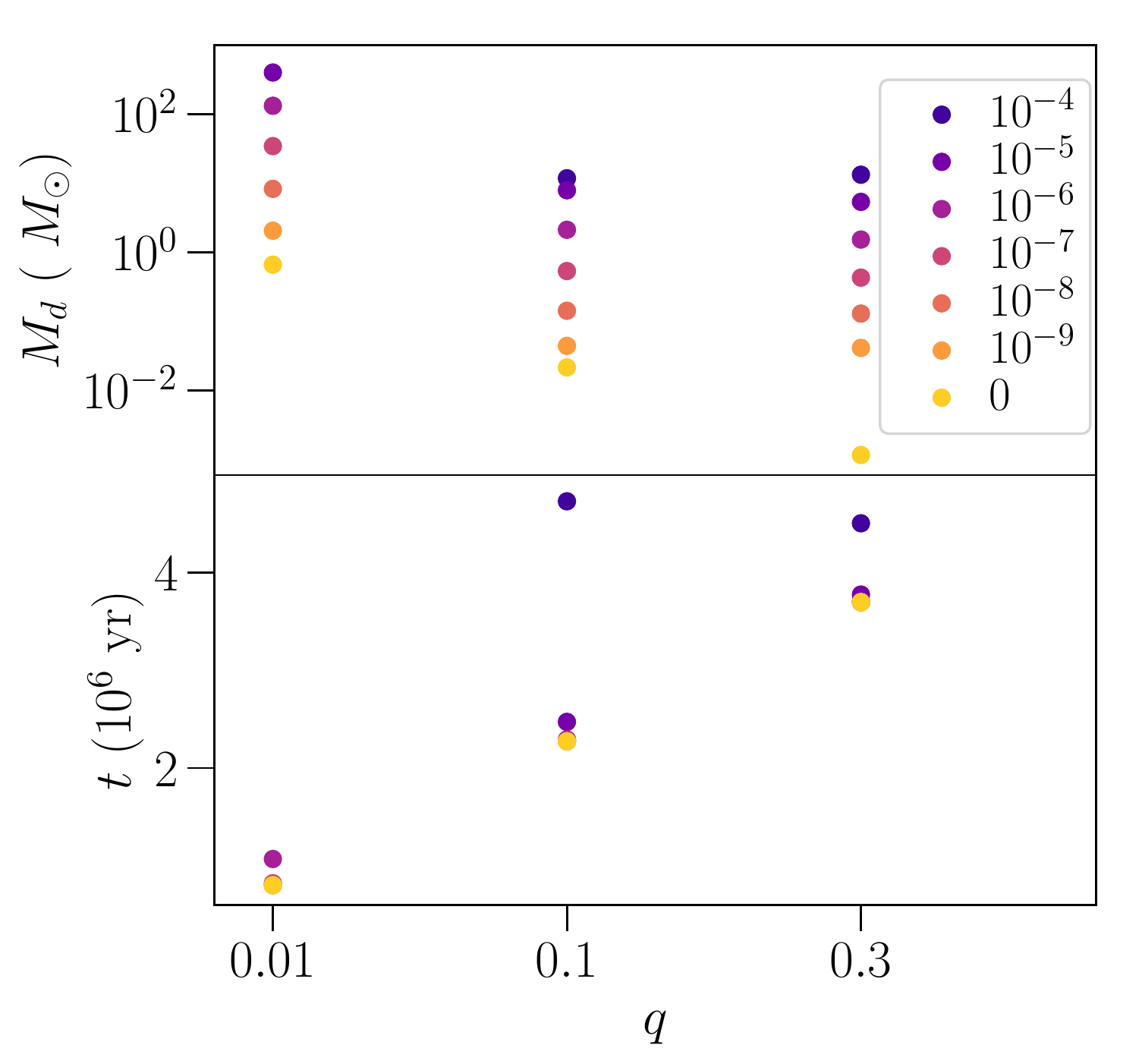}
\caption[Residual mass and time to first decoupling for the first 21 simulations]{Properties of the first 21 simulations. Top panel: Residual mass in the inner disc around the first decoupling, when the binary separation is $a = 10^2 \ {\rm R_S}$. Bottom panel: Time elapsed between a starting binary separation of $a = 10^4 \ {\rm R_S}$ until it reaches $a = 10^2 \ {\rm R_S}$. Both panels are as a function of the mass ratio $q$ and the different values of $\gamma$. While the residual mass changes considerably between simulations, the migration times are similar for each case. Given a fixed mass ratio, as we reduce $\gamma$, each property tends to converge to a value for the case when there is no inflow.}
\label{f:myt}
\end{figure}

The top panel of Fig. \ref{f:myt} presents the residual masses. For each $q$, we find that for smaller $\gamma$, less material is left in the inner disc, as naively expected. However, the difference between the cases is further affected by the position of the gas added in the inner disc \citep{artymowicz94} and how much gas is deposited into it \citep{farris14} Also, removing material from the circumbinary disc changes the surface density distribution itself, affecting the available gas in the vicinity of the secondary. This in turn can change the migration velocity and the inflow rate across the cavity.
Overall, we find that the dependence of the residual mass on $\gamma$ is much weaker than linear. Furthermore, when we shut down the inflow through the cavity entirely ($\gamma = 0$), a fraction of the initial mass of the inner disc still survives, as already found in models that did not allow the gas to flow across the gap \citep{armitage02, chang10, tazzari15}. 

The bottom panel of the same figure shows the time it takes for each system to go from $a = 10^4 \ {\rm R_S}$ to $a = 10^2 \ {\rm R_S}$. We selected $a = 10^4 \ {\rm R_S}$ as a starting separation for this analysis because at this point the shape of the initial condition should long ago have disappeared, making it easier to compare with previous works. The total time that the system needs to reach the final separation does not change much either with the mass ratio $q$ or with the mass influx rate $\gamma$ -- for all our simulations it is close to $3 \times 10^6$ yr, which happens to be very similar to the values found in \citet{tang17}, and also of the same order as the viscous timescale. It is reassuring that our simple torque implementation does not give merger timescales that are wildly different from those found in hydrodynamical simulations, despite the different nature of the torques.

For a given mass ratio, the similarity between the cases with $\gamma < 10^{- 5}$ can be understood as the result of a self--regulation: if material crosses the gap more rapidly, the surface density in the inner disc increases, which adds angular momentum to the binary, and slows down the migration. On the other hand, a higher surface density implies a higher viscosity, which increases the accretion onto the primary BH and reduces the total mass in the inner disc. Finally, in the circumbinary disc, removing material more rapidly produces an enhanced density gradient, which causes the region to re-fill more rapidly, and increases its tidal effects.

For a given $\gamma$, increasing $q$ increases the time needed to reach $a = 10^2 \ {\rm R_S}$. This could be produced by a series of factors: the role of the tidal torque, which becomes stronger for higher mass ratio \citep[but the smoothing becomes relevant, see][]{tazzari15}, the mass of the secondary in the back reaction term in Eq. \ref{eq:dadt}, or the position of the circumbinary disc inner edge. After considering the first two factors, by examining the tidal term in Eq.~\ref{eq:dadt}, we see that the velocity of the secondary increases as $q$ increases, which is the opposite of the trend we found in our simulations. We conclude that the reason for this slow-down comes from the position of the inner boundary of the circumbinary disc: while for $q = 0.01$, $r_{\rm edge} = 1.2 a$, for $q = 0.1$, $r_{\rm edge} = 1.7 a$, and for $q = 0.3$, $r_{\rm edge} = 2.5 a$. Adopting for simplicity the surface density profile for a steady state disc with $\Sigma \propto r^{-3 / 5}$ \citep{kocsis12a}, and using Eq.~\ref{eq:dadt} without the GW contribution, we find that the ratio between the migration velocities are $\dot{a}_{0.01} / \dot{a}_{0.1} = 5.32$ and $\dot{a}_{0.1} / \dot{a}_{0.3} = 3.84$, which could explain the slower gas-driven inspiral we found for larger $q$.

As explained above, $\gamma = 10^{- 4}$ produces an inflow rate similar to a steady-state accretion without the secondary when $q = 0.1$. In this case, the material added to the inner disc doubles the time the binary needs to coalesce, compared with the lower $\gamma$ cases. For $q = 0.3$ this $\gamma$ makes a smaller difference, while for $q = 0.01$ it prevents the binary to merge at all, by driving it outwards.

We stop our simulations when the binary separation becomes $a = 20 \ {\rm R_S}$ because around this point we expect the second decoupling to occur. The inner disc should subsequently become thick, with the secondary ploughing through it. For a more detailed discussion of this last stage of binary evolution, see \citet{fontecilla17}.

\subsection{Is the system in steady state?} \label{sec:steady}

\begin{figure}
\includegraphics[width = 0.93\columnwidth]{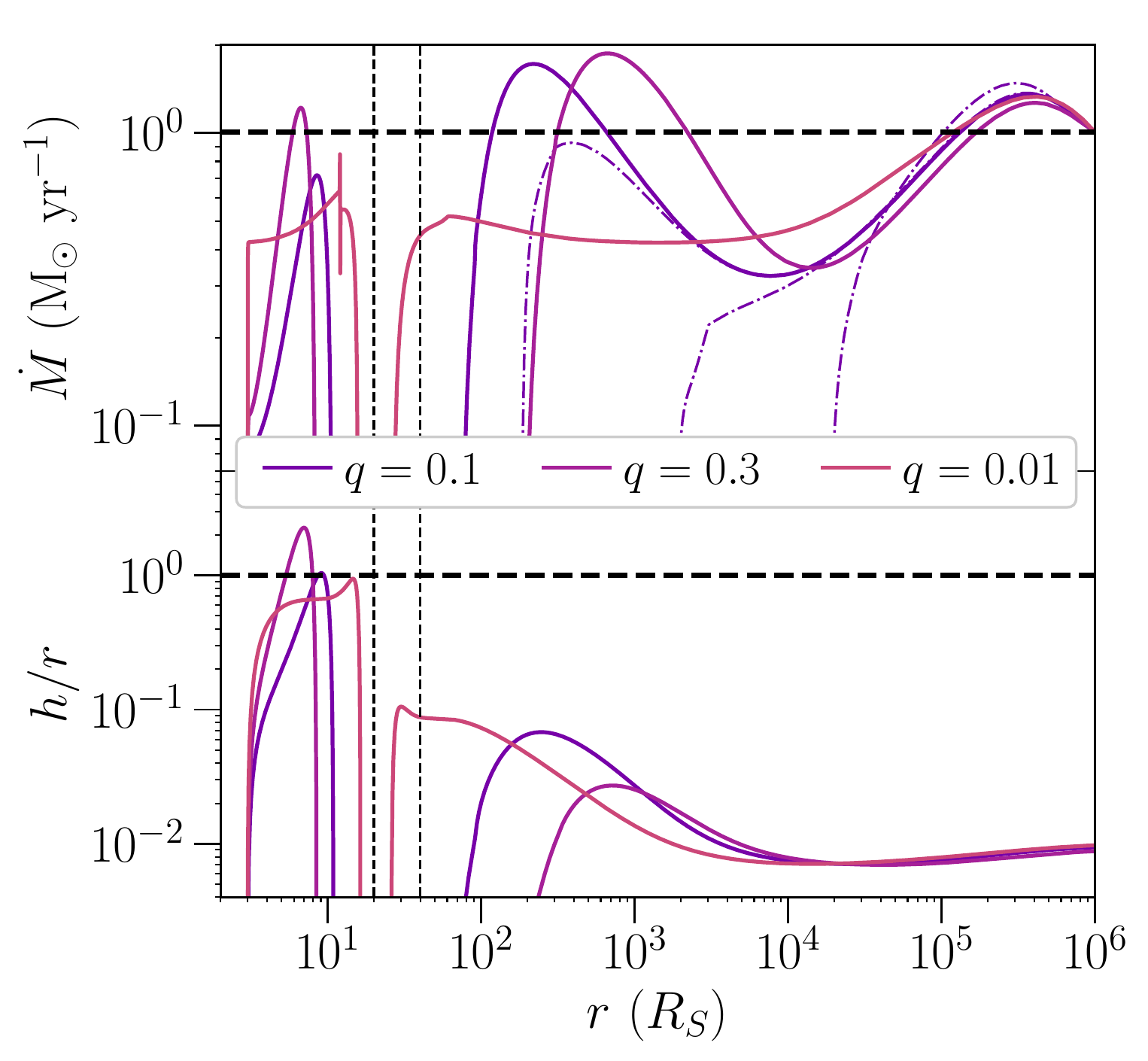}
\caption[Accretion rate and scale height of the discs when the binary separation is $a = 20 \ {\rm R_S}$, as a function of the radius]{Accretion rate (top) and scale height (bottom) of the discs as a function of radius at the binary separation of $a = 20 \ {\rm R_S}$. Results are shown for $\gamma = 10^{-5}$ and for three different mass ratios $q$ as labeled. Dashed curves show the three preceding snapshots at $10^2, 10^3$, and $10^4 R_S$ of the simulation with $q = 0.1$. The radial dependence of the accretion rate is produced by the binary's history of migration, and shows that the disc has not established steady state, even when the binary is close to merger. The scale height in the bottom panel demonstrates that the inner disc becomes thick at this close binary separation, as discussed in \citet{fontecilla17}.}
\label{f:mdot}
\end{figure}

We next address the question of whether the system follows a series of steady state solutions, as often assumed in previous studies \citep[e.g.][]{liu10, kocsis12b, rafikov16}. In the top panel of Fig. \ref{f:mdot}, we show the accretion rate through the disc at the last snapshot of our simulations, for every mass ratio $q$ with $\gamma = 10^{- 5}$. The thick horizontal dashed line indicates the external accretion rate $\dot{M}_{\rm ext}= 1 M_\odot {\rm yr^{-1}}$. A higher or lower value indicates that a pile-up or a depletion is occurring in that annulus of the disc. Because this snapshot is right before the merger, if the steady state assumption were fully correct, the system should have had enough time to smooth out those variations and establish a constant $\dot{M}$ independent of radius. The coloured dashed curves show the three preceding snapshots at $10^2, 10^3$, and $10^4~{\rm R_S}$ of the simulation with $q = 0.1$. As the figure shows, the system is always far away from a steady-state configuration. At the external boundary we have imposed a constant accretion rate as a boundary condition. Going inward, the accetion rate profile can be understood as follows. For $r \gtrsim 10^5 \ {\rm R_S}$ the enhancement in the accretion rate is mostly produced by the viscous torques while the simulation adjusts the boundaries of the accretion discs after the sharp initial condition. In the range $10^4 \ {\rm R_S} \lesssim r \lesssim 10^5 \ {\rm R_S}$ this accumulation disappears. The migration velocity of the binary was fast enough to make the material unable to pile up at the inner boundary. Also, the viscous time of the gas added as the external boundary condition is longer than the merger time of the system, so this gas does not reach the inner region of the circumbinary disc during our simulation. In the range $10^2 \ {\rm R_S} \lesssim r \lesssim 10^4 \ {\rm R_S}$, a steady increment in the accretion rate is produced by the slow-down of the binary in the cases with $q \geq 0.1$, while in the case with $q = 0.01$ this barely happens. In this range, the tidal wall formed near the inner edge of the circumbinary disc stays longer at each radius, increasing the surface density and the accretion rate outside the range of the tidal torque.
\footnote{To see this, consider a disc at two different times $t_2 > t_1$, where $\Sigma_{t_2} > \Sigma_{t_1}$. We assume the discs have a radial dependence of the form $\Sigma_{t_i} \propto r^{d_i}$, where $d_2 < d_1 < 0$ and $d_2 = d_1 - |\Delta d|$ ($\Delta d \ll 1$). Since $\dot{M} \propto - r \Sigma v_r$ and $ v_r \propto - \Sigma^{- 1} r^{- 1 / 2} \partial (\nu \Sigma r^{1 / 2}) / \partial r$, if we use the equations outlined in section \ref{sec:visc}, considering a gas pressure dominated region without tidal dissipation \citep[i.e. outside $r_\Lambda$, see][]{rafikov16}, then $\nu \propto \Sigma^{2 / 3} r$ and $ \dot{M} \propto r^{1 / 2} \partial (\Sigma^{5 / 3} r^{3 / 2}) / \partial r$. With these simplifications, the ratio between the accretion rates becomes:
\begin{equation}
\frac{\dot{M}_{t_2}}{\dot{M}_{t_1}} = \left(\frac{\Sigma_{t_2}}{\Sigma_{t_1}}\right)^{5 / 3} \left(1 - \Delta d \frac{30}{10 d_1 + 9}\right),
\end{equation}
which for $d_1 < - 0.9$ is alway greater than unity, proving our assertion.}
When $r \sim 80 \ {\rm R_S}$ for $q=0.1$ and at $r \sim 220 {\rm R_S}$ for $q = 0.3$, the accretion rate peaks and then falls to zero, showing the decoupling between the binary and the circumbinary disc, since viscosity moves the gas slower than the binary migrates. In the case with $q = 0.01$, the circumbinary disc has not fully decoupled yet, and its inner boundary reaches a distance of less than $2 R_h$ from the binary orbit. This is the closest the disc gets to the secondary in our simulations, in difference from the ``overflow'' scenario described in \citet[][see discussion below]{kocsis12b}. Near the outer boundary of the inner disc, the seconday's torques have squeezed the material inward and enhanced the accretion rate and the thickness of the inner disc. The bottom panel of the same figure shows the scale height: while it is below unity, the tidal effect can push the material outside the binary orbit; but when it becomes of order unity, the thin disc approximation breaks down.

\subsection{Sensitivity to model parameters}

\begin{figure} 
\includegraphics[width = 0.93\columnwidth]{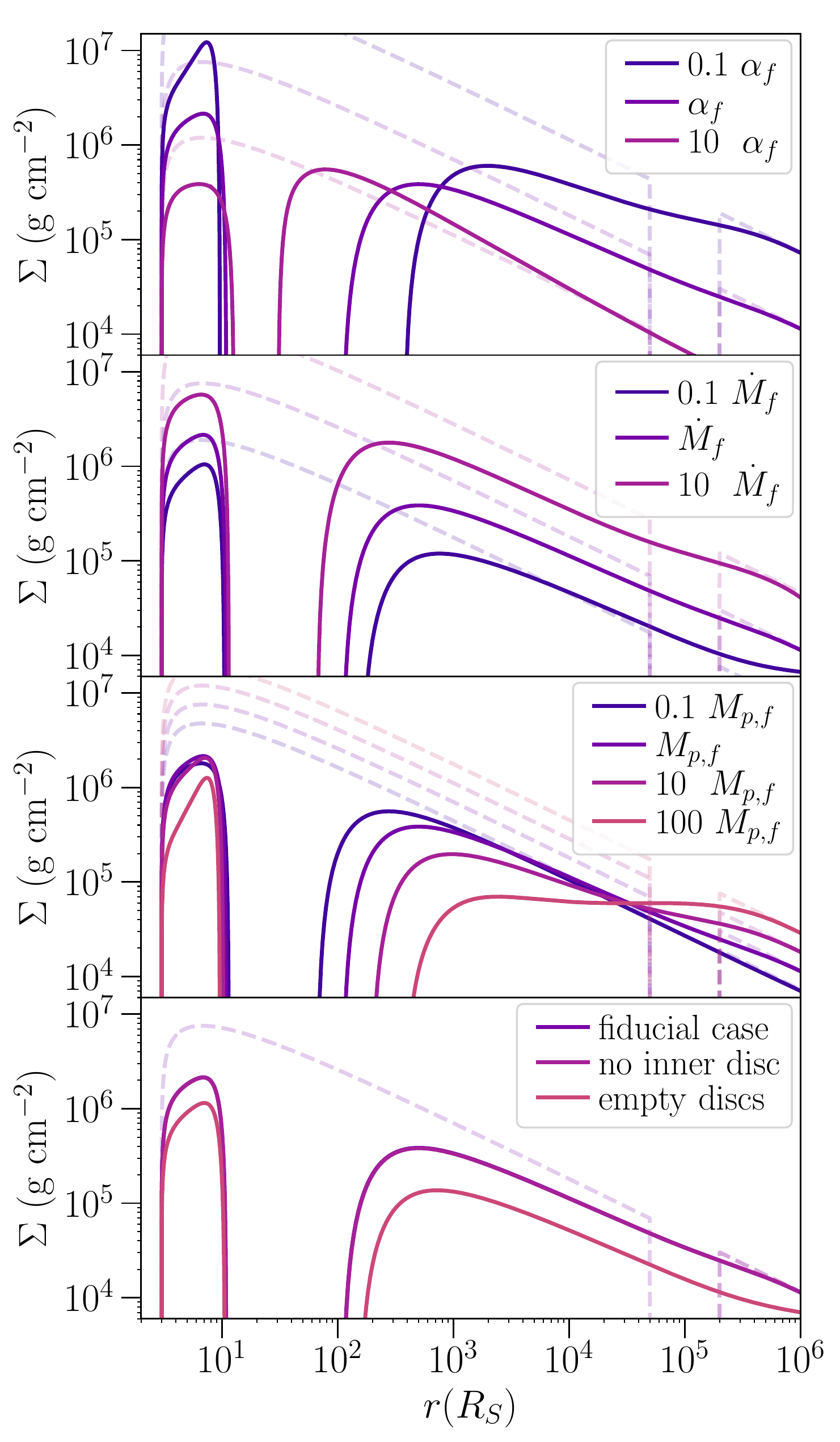}
\caption{Different tests to see how the surface density at the last snapshot ($a = 20 \ {\rm R_S}$) changes with system parameters. The top panel shows simulations with different $\alpha$ viscosity parameters (runs 22 \& 23; see Table~\ref{tab:params} for more details). The second panel shows the results when we change the initial (and boundary) accretion rate (runs 24 \& 25). In the third panel, we change the mass of the primary black hole (runs 26-28) while in the bottom panel, we remove one or both discs from the initial condition (runs 29 \& 30).} 
\label{f:per} 
\end{figure}

In order to assess the sensitivity of our results to the model parameters, we ran several additional simulations with a fixed mass ratio $q = 0.1$ and $\gamma = 10^{- 4}$, but changing one or more parameters of our system, as listed in Table \ref{tab:params}. Fig.~\ref{f:per} shows the surface density profile in each case, at a binary separation $a = 20 \ {\rm R_S}$.

Increasing the strength of viscosity (controlled by $\alpha$, top panel) decreases the initial surface density and enhances the material's inflow speed, which in turn reduces the time the binary needs to merge and produces a less depleted circumbinary disc. On the other hand, increasing the initial and external accretion rates (second panel) increases the initial surface density and reduces the time the binary needs to merge. Still, the overall shape of the surface density, both in the inner and in the circumbinary disc, are overall similar. 

When we increase the primary black hole mass (third panel), the initial surface density, the external accretion rate and the amount of material that crosses the cavity all increase. Since the Schwarzschild radius also depends on the primary mass, and our initial separation is a fixed amount of $\rm{R_S}$, the overall effect is a longer time to merger.

Finally, in the simulation without the inner disc, but allowing the inflow through the cavity ($\gamma = 10^{- 4}$, bottom panel), the final surface density is identical to our fiducial case, but it takes less time to merge. If we remove both discs and allow the external accretion rate to fill the system, it takes ten times longer than in the fiducial case to reach $a = 20 \ {\rm R_S}$. At this moment, the surface density is still around $2-3$ times smaller than in the fiducial case, reinforcing our conclusion that the system never reaches a true steady state in its evolution.

\begin{figure} 
\includegraphics[width = 0.93\columnwidth]{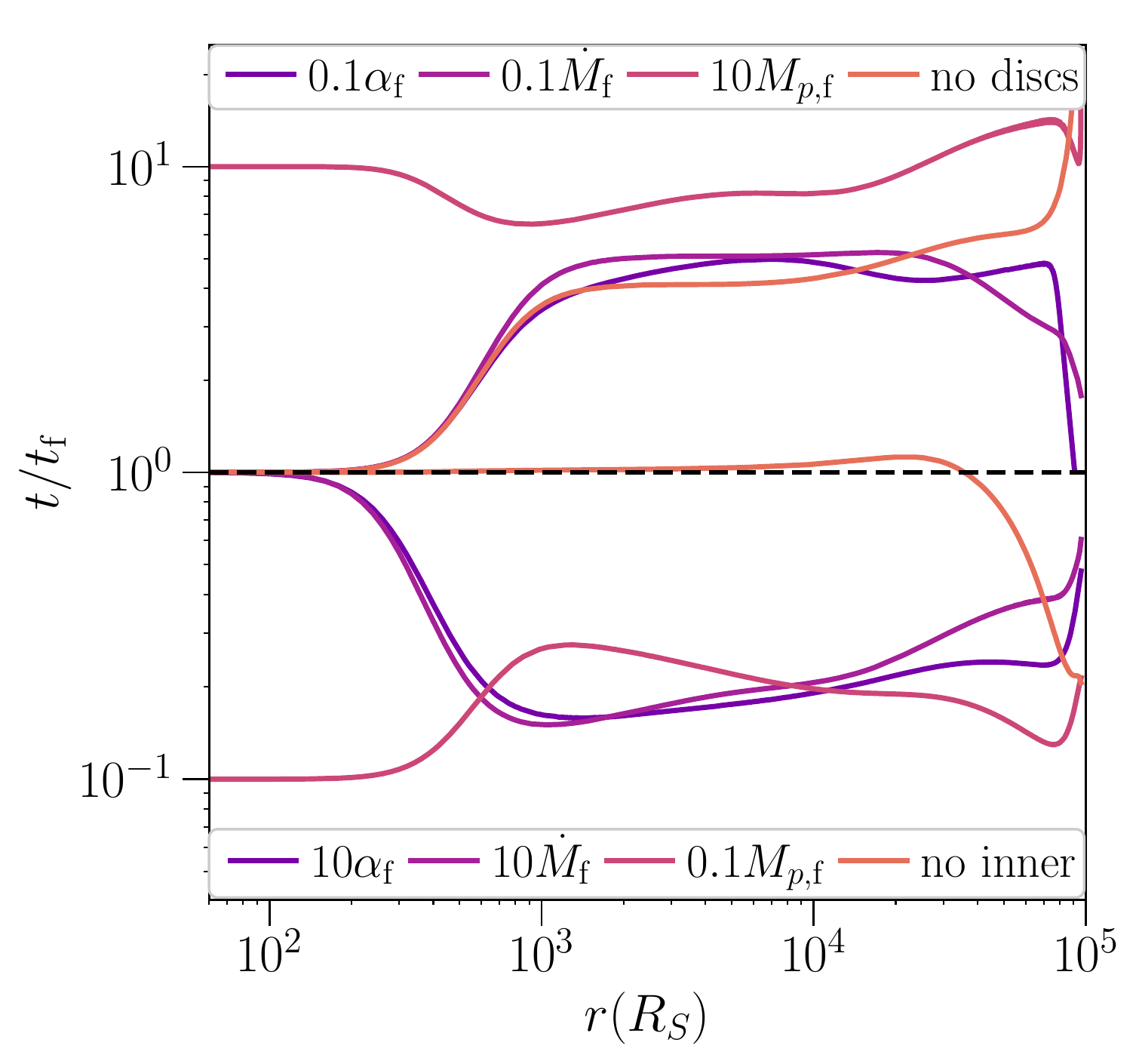}
\caption{Same simulations as in Fig.~\ref{f:per}, but here we show the ratio between the binary's residence time in each case to the fiducial run. As expected, changing the primary mass changes the GW timescale, and it dominates from $a = 10^2 \ {\rm R_S}$ inwards.} \label{f:per2} 
\end{figure}

Fig. \ref{f:per2} shows the ratio between the residence time of each test simulation and our fiducial case. Having no initial discs, reducing $\alpha$ or $\dot{M}$, have similar consequences in this ratio, increasing the residence time by a factor of around four, up to the point where GW dominates. Similarly, increasing $\alpha$ or $\dot{M}$ reduces the residence time by almost an order of magnitude. The simulation with only the circumbinary disc at the beginning becomes indistinguishable from the fiducial case for $a < 10^4 \ {\rm R_S}$. Finally, changing $M_p$ results in an overall scaling of the residence time that can be seen at the end of the simulation.

\section{Comparison with previous works} \label{sec:pw}

In this section we contrast our results with previous studies of the long-term evolution of SMBHB systems. First, we compare our work with analytical estimates that assumed steady-state, and then with other numerical simulations where different conditions were implemented.

\subsection{Previous analytical works}

The first and most relevant work to compare our results with is \citet[][hereafter K12]{kocsis12b}, where the disc properties are obtained analytically, using a similar set of equations, except assuming that the accretion discs go through a series of steady-state solutions as the binary evolve. Since in our simulations we dropped the steady-state approximation and let the binary separation shrink over time by both the influence of the discs and GWs, this comparison can give us insight into whether the steady state assumption is justified.

As we saw in Fig. \ref{f:rest}, in the range between $10^2 \ {\rm R_S} \lesssim a \lesssim 10^4 \ {\rm R_S}$ the inflow timescale of the circumbinary disc is shorter than the binary's residence time. While this is a necessary condition for steady state, it is not sufficient, as we already discussed in section \ref{sec:steady} above (see Fig. \ref{f:mdot} in particular).

There are two other important differences between K12's model and the present study. First, K12 did not include the GW contribution in their calculations, since when the GWs dominate, the steady-state assumption is clearly invalidated. Second, K12 treats the gas inflow at the secondary differently. In particular, they assume that the gas near the secondary flows inward at the constant {\em accretion rate} $\dot{M}_{\rm{ext}}$, and demand that the inward {\em radial velocity} of the gas at the inner edge of the circumbinary disk (at $r\approx 2a$) is $\approx$ twice the migration speed of the secondary. This envisions a steady-state configuration, in which the gas just outside the secondary's orbit follows the inward-migrating secondary, preserving the approximately 2:1 ratio of the size of the gap and the binary separation. Depending on the values of the system parameters, solutions with this prescription are found with or without an inner disc. In summary, K12's prescription has no free parameters and describes a steady-state, in difference from ours, in which we move mass inward across the gap at a rate parameterized by the new parameter $\gamma$, and require no steady state. Since K12 excludes GWs, we can only compare our results with their snapshots at the relatively large binary separation of $a = 500 \ {\rm R_S}$. To make a direct comparison, we ran four new simulations with the parameters in K12, $\dot{M}_{\rm{ext}} / \dot{M}_{\rm{Edd}} = 0.1$, $q = \{0.1, 0.01\}$ and $\gamma = \{10^{- 5}, 0\}$ while the rest of the parameters are the same as our fiducial case. As our test with $0.1 \dot{M}_f$ demonstrated, when we vary the accretion rate through the disc, the final shape of the surface density is similar, and the only difference is in its overall scaling. At $a = 500 \ {\rm R_S}$, K12 found a solution with a completely depleted inner disc (see their Fig.~2), and a strong pile-up in the outer disc. The increase in the circumbinary disc surface density is a factor of $3$ over the initial condition for $q = 0.1$ and $2.1$ for $q = 0.01$, with no material inside the secondary orbit.

These results most closely corresponds to our case without any inflow across the gap ($\gamma = 0$), for which we found a more modest surface density enhancement of a factor of $1.5$ for $q = 0.1$, and no enhancement at all for $q = 0.01$. While in the former case we also obtain a gap that extends all the way down to the innermost stable circular orbit (ISCO), in the latter we found an inner disc, albeit with a very low surface density ($250$ times smaller than the initial condition). Finally, the thickness of the circumbinary disc in our case is three times smaller than in their results, while our $\beta$ parameter is more than five times higher. At this separation K12 found, for $q = 0.01$, a binary radial velocity $da/dt \sim 200 \ \rm{cm / s}$ and a disc velocity (measured at $r = 2a$) two times faster. Instead, in our simulation with the same mass ratio and $\gamma = 0$, our binary radial velocity is $da/dt \sim 60 \ \rm{ cm /s}$ and the disc velocity is four times faster. Those values are consistent with our surface density being half theirs. Since we run a numerical simulation from farther away and let the system reach this point solving the equations self-consistently, we consider our results to be a more precise estimation of the disc properties at this binary separation.

We note further that a better definition for a ``pile-up" is not the overall increase in the surface density, but rather the change in the slope of the discs at different times. The surface density of our discs are generally smaller than K12's because our system never reaches a steady-state configuration. If we were to fix the binary separation and let the discs evolve, the disc masses would increase. To illustrate this, we re-normalized our model output to match their surface density at $ r = 5 \times 10^3 \ {\rm R_S}$, and found a slope very similar to K12's at $a = 500 \ {\rm R_S}$.

To further test K12 set-up, we run a simulation with $\gamma = 10^{- 5}$. Our result resembles the case on the left panels in our Fig. \ref{f:data}, where no pile up is found and an inner disc is maintained over all the simulation. An smaller $\gamma$ could reproduce the circumbinary disc pile-up, with a depletion and subsequent refill of the inner disc, depending on the binary separation.

The fact that K12's surface density is enhanced by a factor or two in comparison with their initial condition (and our model), also partly explains why their residence time is ($\approx$ four times) shorter than ours.

To explore the residence times further, we next focus on the results of \citet[][hereafter H09]{haiman09}. Unlike K12 or the present work, H09 did not incorporate the binary torques directly into the disc structure equations or the resulting migration rate. Instead, they used the density and viscosity profiles of the usual Shakura-Sunyaev discs, modified by the presence of the secondary as detailed in prior semi-analytic studies \citep{syer95, ivanov99}. They assumed that the secondary's migration is tied directly to the viscous time in the modified disc, and presented a dedicated analytical study of the resulting residence times.

To make a meaningful comparison, we run a new simulation with a parameter combination included in H09, i.e, $\alpha = 0.3$, $\dot{M}_{\rm{ext}} / \dot{M}_{\rm{Edd}} = 0.1$, M = $10^7 M_\odot$ and $q = 0.01$. The main difference is that we solve the system self-consistently, allowing the disc structure to be modified in a different way, depending on the migration rate of the secondary.
 
The residence time we obtained is similar to H09 at both ends, i.e. when the GWs dominate ($a \lesssim 50 \ {\rm R_S}$) and when the binary separation is bigger than $a \gtrsim 2 \times 10^4 \ {\rm R_S}$, where the type II migration should be controlled by the circumbinary disc (with disc mass exceeding the secondary's mass). In-between these separations, our residence times deviate from H09's results for an unsteady $\beta$ disc (based on \citealt{ivanov99}) with the largest difference at $400 \ {\rm R_S}$, when ours is two times slower. Then, at $a \sim 200 \ {\rm R_S}$, our solution becomes dominated by GW, decreasing until both results become similar again at $a = 50 \ {\rm R_S}$.

One explanation for this difference could be that in our work we made the direct calculation of the tidal torque, both from the inner and the circumbinary disc, whereas the inner disc is neglected in H09. The presence of an inner disc will add angular momentum to the binary and will slow down the migration. Moreover, as the binary shrinks, the interaction region defined by the Lindblad resonances also shrinks. This effect is stronger for the circumbinary disc, making the contribution of the inner disc more significant at later times. The difference can also be partially produced by the tidal torque used. In their fiducial model, H09 follows the self-similar disc evolution in \citet{syer95}, which uses a tidal torque:
\begin{equation}
\Lambda' = q^2 \Omega^2_s a^2 \bigg(\dfrac{a}{\Delta}\bigg)^4 = \frac{2 r}{a f} \Lambda \sim (2 - 8) \times 10^2 \Lambda,
\end{equation}
in the region of tidal interaction. Therefore, the tidal effect is stronger in their work.

In summary, Table \ref{tab:restime} shows our residence times for the simulations mimicking K12 and H09, along with their results for a separation of $a = 400 \ {\rm R_S}$ and $a = 10^4 \ {\rm R_S}$. We see that, while for $a = 10^4 \ {\rm R_S}$ the analytical and numerical results are similar, at closer separations the absence of an enhancement in the circumbinary disc surface density and the presence of an inner disc made our residence times longer. 
\begin{table}
\centering
	\caption{Residence times in our simulations, compared to corresponding semi-analytic results from \citet{kocsis12b} and \citet{haiman09} with similar parameters ($q = 0.01$ and $\gamma = 0$).}
	\label{tab:restime}
  \begin{tabular*}{0.7\columnwidth}{lcc} 
     & $a = 400 \ {\rm R_S}$ & $a = 10^4 \ {\rm R_S}$ \\ 
    \\[-2ex]
		\hline
    \\[-2ex]
    \citeauthor{kocsis12b} & $1.5 \times 10^5$ yr & $3.6 \times 10^6$ yr\\
    Run 36 & $~~7 \times 10^5$ yr & $4.6 \times 10^6$ yr\\
    \citeauthor{haiman09} & $~~2 \times 10^5$ yr & $1.5 \times 10^6$ yr\\
    Run 38 & $~~4 \times 10^5$ yr & $2.1 \times 10^6$ yr\\
    \hline     
	\end{tabular*}
\end{table}

\begin{figure}
\includegraphics[width = 0.93\columnwidth]{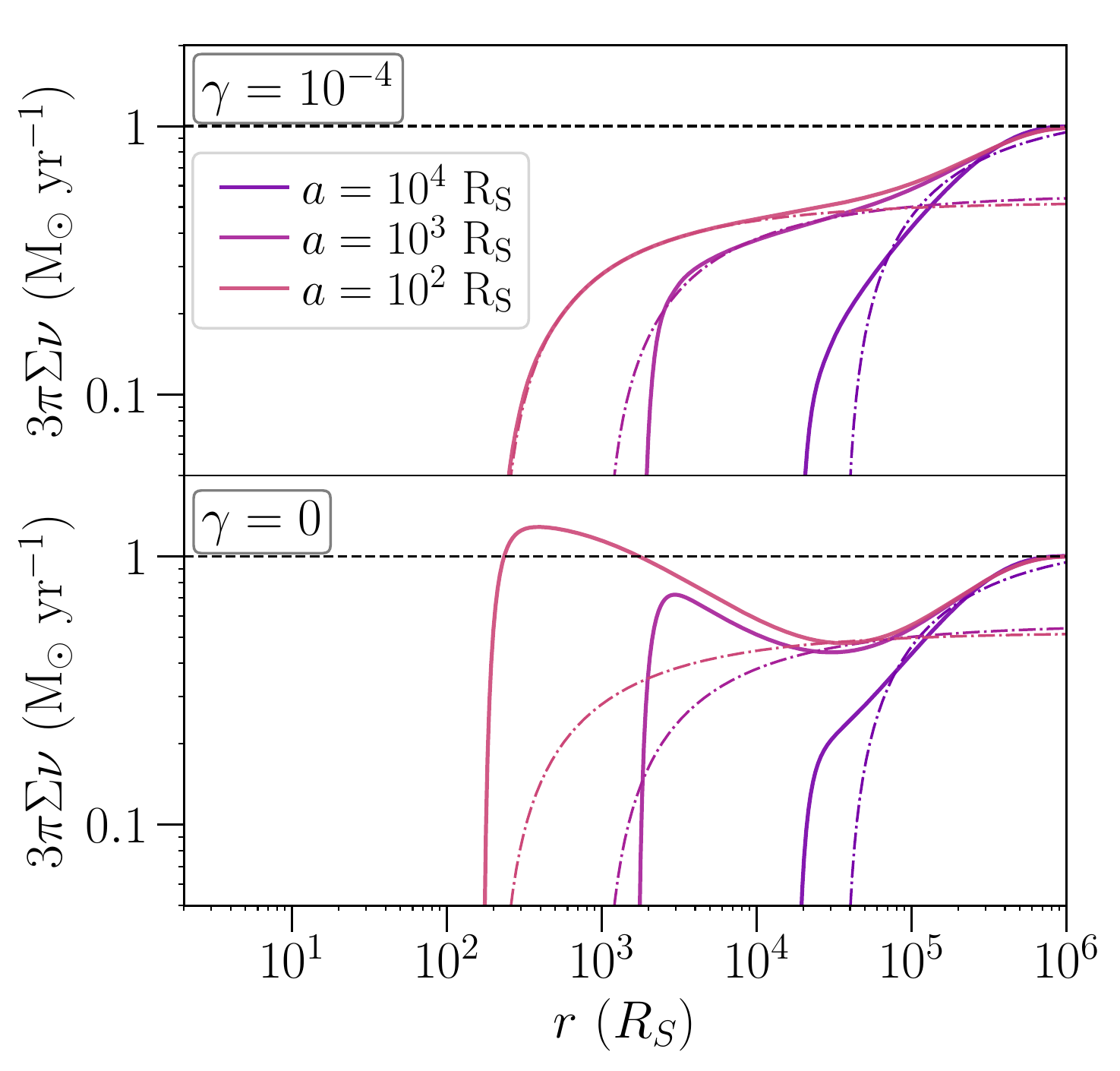}
\caption{The radial dependence of the accretion rate in our simulations, compared to the semi-analytic results of \citet{rafikov16} which are expected to be valid far away from the binary. The solid curves show the simulation results for three different binary separations, as labeled, while the dashed curves show analytic fits to the results of \citet{rafikov16} from Eq. \ref{eq:rafikov} for $\gamma = 10^{- 4}$. In the top panel, the analytic profiles follow our results with $\gamma = 10^{- 4}$ closely beyond 2-3 times secondary's orbit. In the bottom panel, the case with no inflow through the cavity ($\gamma = 0.0$) shows a clear pile up over time, causing much stronger deviations at small radii.}
\label{f:rafikov} 
\end{figure}

Finally, we compare our results with the work of \citet{rafikov16}, which presented analytical estimates of the circumbinary disc structure, assuming steady-state evolution. Note that \citet{rafikov16} did not include the binary torque directly in the evolutionary equations. The torques were instead added as an additional flux of angular momentum, injected at $r=0$. The solutions should nevertheless be accurate far outside the binary-disc interaction region, but only if steady state is established. Following \citet{tang17}, we took Rafikov's equations (3), (11) and (20), and write:
\begin{equation}
3 \pi \nu \Sigma = {\dot{M}(r_b)} - \left.\frac{q M_p}{2 (1 + q) \sqrt{a r}} \frac{d a}{d t}\right|_c = A - \frac{B}{\sqrt{r}}. 
\label{eq:rafikov} 
\end{equation} 
Here, $\dot{M}(r_b)$ is a constant accretion rate at a radius $r_b \gg a$ far away from the secondary, and $d a / d t |_c$ is the circumbinary disc contribution to the binary migration. Fig. \ref{f:rafikov} shows the best fit for $A$ and $B$ in each snapshot, at different binary separations, for our simulations with $\gamma = 10^{- 4}$. The bottom panel shows our numerical results for $\gamma = 0$, compared to the same analytic fits as in the top panel. 

In the top panel, the analytic profiles follow our results closely, in each case, beyond 2-3 times secondary's orbit. In the bottom panel, the case with no inflow through the cavity shows a clear pile-up over time, causing much stronger deviations extending to much larger radii. As discussed above, $\gamma = 10^{- 4}$ is the value at which our mass flow rate across the cavity matches the value expected for steady state accretion, while $\gamma = 0$ turns off this cross-cavity flow. Since the inflow timescale is shorter than the migration time, material piles up in the inner region of the circumbinary disc. Furthermore, in this case, the profiles deviate from the analytic expectations much farther out. Rather then always remaining approximately correct out to 2-3 times the current binary separation, the analytic solutions are now accurate only beyond a few $\times 10^4 \ {\rm R_S}$, even at late times and much smaller binary separations. Because the system falls increasingly out of steady state, this critical radius (beyond which the analytic solution is valid) now remains tied to the initial conditions, rather than to the current binary separation. 

\subsection{Previous numerical simulations}
In \citet{lodato09}, the authors study a similar setup as our own, with the difference that their initial condition for the accretion discs has a fixed total mass and they do not have any inflow of material from the outer boundary.

To compare our results to their work and to see how our initial and boundary conditions change the time to merger, we ran a new simulation with their parameters (No. 39 in Table \ref{tab:params}). The cutoff from \citet{tazzari15}, also implemented here, prevents the material outside the interaction region to directly influence the migration of the binary. To compute the mass of the interacting disc, we integrate from ISCO out to a radius where the tidal contribution to the total radial velocity of the gas is less than one percent.

\citet{lodato09} considered an initial binary separation $a_0 = 0.01 {\rm pc} \sim 10^3 \ {\rm R_S}$ and concluded that, for an initial disc with half the mass of the secondary, the merger happens after $4.8$ Myr. Our interacting disc mass at the same binary separation is around $0.4 M_s$, and in our simulation the system needs $5.8$ Myr to merge, very close to their results. At the moment of the first decoupling, \citet{lodato09} found that an inner disc survives even for the case with an initial disc mass ten times smaller than the secondary. In our simulation, the inner disc is completely accreted, which we suspect it due to the different initial condition used (since $\gamma = 0$).

Finally, we compare our residual mass after the first decoupling with that obtained by \citet{tazzari15}. Compared with our simulation without inflow, their result is half the one we obtained, and even smaller for the cases where $\gamma \neq 0$. 
\footnote{\citet{chang10} had found a much smaller residual mass, due to the overestimate of the classical formula for the tidal torque (see eqs. \ref{eq:smooth} and \ref{eq:smooth2}, and \citealt{tazzari15}).} The main differences between our works are:(i) we have an inflow of material from the circumbinary disc, controlled by $\gamma$ and (ii) our external accretion rate is ten times higher than theirs. In the simulation with no inflow, the external accretion rate has no effect in the inner disc since material cannot cross the cavity. If we compare our inner disc mass at $a = 10^4 \ {\rm R_S}$ (their initial condition), ours is three times larger, which could explain the final difference.

In summary, we emphasize that the key novel factor in our work is the implementation of the inflow through the cavity. This ingredient changes the final state of the binary at the first decoupling, as we can see from the previous comparisons. While we do not know exactly which value of $\gamma$ is the `right' one, or if it changes over time, our results show how important this inflow rate is for setting the residual mass, while we find that it does not greatly influence the residence time of the binary.

\section{Summary and Conclusions} \label{sec:con}

In this work we developed 1D models to study the long-term, coupled evolution of a SMBH binary + gas disc system, solving the equations for the evolution of the surface density and temperature, and the secondary BH's migration in a self-consistent way. The non-axisymmetric distortions in the background discs, the resulting torques, and the rate at which gas can cross the gap created by the secondary, cannot be computed within our model. We instead incorporate these effects by prescriptions motivated by 2D simulations. In particular, the torques are implemented via a fitting formula calibrated to simulations \citep{armitage02}. We further introduced the parameter $\gamma$, which specifies that in the absence of viscosity, the material from the circumbinary disc would be transported to the inner disc in $\sim \gamma^{- 1}$ binary orbits, due to the secondary BH's gravitational torques and shocks. We implemented and tested different values of this parameter. While a 1D simulation is unable to capture the full range of physical processes at hand, it remains the only practical way to study the coupled long-term evolution of such systems.

The main results of our study are the residual mass in the inner disc near the end of the merger (when the binary separation is $a = 10^2 \ {\rm R_S}$), the time for the BHs to merge, or, in practice, to reach the GW-driven regime (starting from $a = 10^4 \ {\rm R_S}$), as well as the time-evolving deviations in the spectrum and luminosity produced by the tidal heating. All of these quantities are functions of the system parameters, including $q$~and~$\gamma$.

The residual mass is very sensitive to system parameters, and ranges between $10^{-2}$ to $10^2 M_\odot$, with a variation by two orders of magnitude at a given mass ratio. In particular, if gas can cross the gap efficiently (large $\gamma$), the large amount of residual gas is a promising source of fuel for a possible EM counterpart to LISA events, as proposed in earlier work \citep{chang10, tazzari15}. By comparison, we find the merger time to be much more robust: the total time for the secondary to migrate from $a = 10^4 \ {\rm R_S}$ to $a = 10^2 \ {\rm R_S}$ is always within a factor~of~$\lesssim 3$ of $2 \times 10^6$ yr. These values are reassuringly similar to those found by directly measuring the total torque in 2D simulations \citet{tang17}, and also of the same order as the viscous timescale. They reinforce the conclusion that a circumbinary disk, fueled at near-Eddington rates, can drive the binary to coalesce well within a Hubble time. 

As expected, when we decrease $\gamma$ the pile-up in the circumbinary disc becomes more pronounced and the inner disc depletes. This can be seen in the SED: the hot region of the circumbinary disc is enhanced while the inner disc diminishes. The time before merger when Bolometric Luminosity peaks also depends on the amount of material that crosses the cavity. We find that for $\gamma \geq 10^{- 5}$ this happens at the ``second decoupling'', while for smaller values, the tidal heat of the circumbinary disc at the ``first decoupling" dominates. 

We also found that in general, the binary + disc system never reaches a genuine steady state. This is most clearly seen in our analysis of the accretion rate profile (\S~\ref{sec:steady}), where the accretion rate remains a function of radius. The entire evolutionary history of the system generally depends on the initial and outer boundary conditions. When we assume the gas crosses the gap inefficiently (e.g. in our runs with $\gamma \leq 10^{- 5}$ and $q = 0.1$), we find that the inner disc is depleted and then re-filled (see Fig. \ref{f:data}), while the bolometric luminosity increases and decreases by an order of magnitude (Fig. \ref{f:bol}). When we assume the gas can cross the gap more efficiently (i.e. for higher values of $\gamma$), this depletion does not occur. In the case with $\gamma = 10^{- 4}$, an approximately constant surface density develops over time, with the magnitude and the slope of the surface density profiles similar between snapshots, but the accretion rate profiles still show that a steady state is not established (\S~\ref{sec:steady}). This non-steady-state behaviour implies that the gas distribution near the time of the merger and close to the BHs depends on the boundary conditions at large radii and at early times. This issue must be studied further in 2- and 3D simulations, and eventually incorporated into the initial conditions of more sophisticated (e.g. 3D GRMHD simulations), designed to follow the last stages of the merger.

\section*{Acknowledgments}
CF and JC acknowledge financial support from CONICYT-Chile through FONDECYT (1141175) and Basal (PFB0609) grants. CF acknowledges financial support from CONICYT-PFCHA/Doctorado Nacional (2017-21171063). CF thanks Columbia University for warm hospitality, as well as financial support through Columbia University's President's Global Innovation Fund, during his visit where this work started. ZH gratefully acknowledges financial support was by NASA through ATP grant NNX15AB19G, ADAP grant NNX17AL82G, and Swift grant 16-SWIFT16-0015, and by the NSF through grant 1715661. JC acknowledges the warm hospitality of MPE, where part of this work was conducted, and funding from the MPG through a Partner Group grant.

\bibliographystyle{mnras}

\bsp
\label{lastpage}
\end{document}